# United Theory of Planet Formation (I): Tandem Regime

Toshikazu Ebisuzaki and Yusuke Imaeda[a]

[a] 2-1 Hirosawa 351-0198, Japan

**Abstract**

The present paper is the first one of a series of papers that present the new united theory of planet formation, which includes magneto-rotational instability and porous aggregation of solid particles in a consistent way. We here describe the "tandem" planet formation regime, in which a solar system like planetary systems are likely to be produced.

We have obtained a steady-state, 1-D model of the accretion disk of a protostar taking into account the magneto-rotational instability (MRI). We find that the disk is divided into an outer turbulent region (OTR), a MRI suppressed region (MSR), and an inner turbulent region (ITR). The outer turbulent region is fully turbulent because of MRI. However, in the range, $r_{\rm out}(= 8 - 60$ AU) from the central star, MRI is suppressed around the midplane of the gas disk and a quiet area without turbulence appears, because the degree of ionization of gas becomes low enough. The disk becomes fully turbulent again in the range $r_{\rm in}(= 0.2 - 1$ AU), which is called the inner turbulent region, because the midplane temperature become high enough ($> 1000$ K) due to gravitational energy release.

Planetesimals are formed through gravitational instability at the outer and inner MRI fronts (the boundaries between the MRI suppressed region (MSR) and the outer and inner turbuent regions), because of the radial concentration of the solid particles. At the outer MRI front, icy particles grow through low-velocity collisions into porous aggregates with low densities (down to $\sim 10^{-5}\,{\rm g\,cm^{-3}}$). They eventually undergo gravitational instability to form icy planetesimals.

On the other hand, rocky particles accumulate at the inner MRI front,

*Email address:* ebisu@postman.riken.jp (Toshikazu Ebisuzaki and Yusuke Imaeda)

since their drift velocities turn outward due to the local maximum in gas pressure. They undergo gravitational instability in a sub-disk of pebbles to form rocky planetesimals at the inner MRI front. They are likely to be volatile-free because of the high temperature (> 1000 K) at this formation site. Such water-free rocky particles may explain the formation of enstatite chondrites, of which the Earth is likely to be primarily composed of.

Our new proposed tandem planet formation regime shows that planetesimals are formed at two distinct sites (outer and inner edges of the MRI suppressed region). The former is likely to be the source of outer gas giants and the latter inner rocky planets. The tandem regime is consistent with the ABEL model, in which the Earth was initially formed as a completely volatile-free planet. The water and other volatile elements came later through the accretion of icy particles by the occasional scatterings in the outer regions.

The tandem regime also explains the gap in the distribution of solid components (2-4 AU), which is necessary to form a"solar-system-like" planetary system, which has a relatively small Mars and a very small mass in the main asteroid belt .

We found that this tandem regime dose not take place when the vertical magnetic field of the disk five times weaker compared with that we assumed in the present paper, since the outer MRI front shift outward beyond 100 AU. This suggests that yet other regimes exists in our united theory. It may explain the variation observed in exsoplanetary systems by variations in magnetic field and probably angular momentum of the parent molecular cloud.

---

## 1. Introduction

Planets are believed to be formed in a proto-planetary gas disk around a newly born star by the gravitational collapse of a dense molecular cloud (e.g., Bouvier et al., 2007). The spectacular growth of solid particles covers 40 orders of magnitude in mass ($10^{-10} - 10^{30}$ g). The standard scenario of planet formation was first systematically described by Safranov (1969) and extended by C. Hayashi and his colleagues (e.g., Hayashi et al., 1985), as well as many researchers (Goldreich and Ward, 1973; Weidenschilling, 1977a,b;

Wetherill and Stewart, 1989). The main stream of the standard model is divided into the following steps:

1. Submicron or micron-sized interstellar dust grains grow into cm-sized pebbles through mutual collisions and resulting chemical bonding and electrostatic forces, and gradually settle to the midplane of the disk to form a sub-disk of densely concentrated pebbles (Weidenschilling, 1977a,b; Nakagawa et al., 1981; Hayashi, 1981; Hayashi et al., 1985).
2. The pebbles further grow into planetesimals by collisional sticking and then undergo gravitational instability (Safranov, 1969; Hayashi, 1972; Goldreich and Ward, 1973; Weidenschilling and Cuzzi, 1993).
3. Gravity of planetesimals, with kilometer or larger sizes, becomes strong enough to retain collision fragments, leading to Moon-sized protoplanets through mutual collisions by gravitational interaction (Nakagawa et al., 1981; Wetherill and Stewart, 1989).
4. The system of the protoplanets evolves as a gravitational N-body system through gravitational interaction and occasional giant impacts among them (Kokubo and Ida, 2012). The protoplanets, which reach the critical mass for trapping gas before the disappearance of gas in the protoplanetary disk, become gas giants (Mizuno, 1980; Pollack et al., 1996), otherwise become terrestrial planets (Chambers and Wetherill, 1998) without thick gas envelopes.

The major assumptions of the standard model are:

**Assumption A:** Turbulence had diminished in the protoplanetary nebula when particle growth began.

**Assumption B:** Planetesimals formed through gravitational instability in a sub-disk of densely concentrated pebbles.

**Assumption C:** Solid particles did not move considerably in radius after their sedimentation.

**Assumption D:** Planetesimals grew through their mutual collisions.

However, all four assumptions are challenged by recent detailed studies. First, **Assumption A** may not be true, since Balbus and Hawley (1991) and Hawley and Balbus (1991) found that Magneto-Rotational Instability can excite turbulence in the accretion disk and probably also in the protoplanetary nebulae (Sano and Miyama, 1999; Sano et al., 2000), if the ionization degree is high enough.

Second, contrary to **Assumption B**, gravitational instability does not take place, since the sink-down motion toward the midplane terminates well before the onset of gravitational instability (Goldreich and Ward, 1973; Weidenschilling, 1977a,b; Cuzzi et al., 1993; Champney et al., 1995). When the density of the solid particles dominates over that of gas, i.e., $\rho_{\rm p} > \rho$, shear turbulence is driven by the velocity difference between the inside (particle dominated) and outside (gas dominated) parts of the sub-disk. The particles tend to rotate slower than Kepler velocity , $v_{\rm K}$, by $\eta v_{\rm K}$ ($\eta \sim 10^{-3}$) due to the pressure gradient in the disk. In such a situation, the gravitational instability can take place only when the density of particles becomes high enough through the radial migration of the particles (Sekiya, 1998; Youdin and Shu, 2002), at least by a factor of four (Brauer et al., 2008).

Third, contrary to **Assumption C**, viable radial motion of solid components is expected in the protoplanetary disk (Whipple, 1972; Adachi et al., 1976; Weidenschilling, 1977a) reported that solid particles move considerably inward, since the gas in a disk moves slightly sub-Keplerian due to a radial pressure gradient. For this reason, the solid particles, which move with near-Keplerian velocity, face a continuous headwind of gas. Hence, the particles lose their angular momentum due to the drag force between gas and dust particles. The range in timescale of inward migration is as short as $10^2 - 10^3$ yrs. The solid particles eventually drift into the inner evaporation zone near the central star and are lost from planetesimal formation (**drift barrier**).

In addition, solid particles are also difficult to grow beyond meter-size because of the disruption by high-velocity impacts (**fragmentation barrier**; Blum et al., 1998; Poppe et al., 1999; Blum and Wurm, 2000). Wada et al. (2009) estimated the critical velocity of $\sim 60\,{\rm m\,s^{-1}}$ for icy particles and $\sim 1\,{\rm m\,s^{-1}}$ for silicate (rocky) particles, based on the theory of ballistic cluster-cluster aggregation (BCCA). On the other hand, the relative velocity of particles of meter-sized bodies in a protostellar disk can be more than $30\,{\rm m\,s^{-1}}$ (Adachi et al., 1976; Weidenschilling, 1977a; Markiewicz et al., 1991; Weidenschilling and Cuzzi, 1993; Dominik et al., 2007; Blum and Wurm, 2008). In addition, considering the minimum-mass solar nebula model of Hayashi (1981), the radial inward velocity of solid particles reaches $50\,{\rm m\,s^{-1}}$, owing to coupling with disk gas (Adachi et al., 1976; Weidenschilling and Cuzzi, 1993; Suyama et al., 2008). Yet, the collision velocity could be even higher due to gas turbulence. For example, strong gas turbulence expressed by the standard $\alpha$ model with $\alpha = 0.01$ leads to velocity up to $\sim 100\,\mathrm{m\,s^{-1}}$.

Furthermore, protoplanets (or gas-giant cores) larger than the mass of the Earth are rapidly removed from the disk due to the outward migration by the gravitational torque from the gas disk (Goldreich and Tremaine, 1979, 1980; Lin and Papaloizou, 1979; Tanaka et al., 2002). The migration timescale is as short as $10^5$ yr for the case of Earth-sized protoplanets.

Finally, the growth through mutual collisions of the planetesimals (**Assumption D**) is found to be too slow for gas and icy planets, if the saturation of runaway growth is taken into account (Wetherill and Stewart, 1989; Ida and Makino, 1992a,b, 1993). The column density of the solid component must be several times higher than the minimum-mass solar nebula (MMSN) to match the timing of gas dissipation around $\sim 1 - 10\,\mathrm{Myr}$ (Pollack et al., 1996).

In summary, the standard model is invalid and alternative models have yet to be proposed until now. Although, many individual ideas are proposed to overcome one of the difficulties, there still is no overarching solution or picture to explain the formation process of planets or planetary systems. Perhaps this is because it is exceedingly challenging for one person to cover the varied yet multiple aspects of physics in order to follow all of the processes involved in particle growth from sub-micron dust to gas giants ranging 15 orders of magnitudes in radius

Some investigators have proposed their ideas to solve problems based on one or two assumptions. In the case of **Assumption A**, for example, MRI can be inactive near the midplane in the region $r < 10$ AU because of a very low degree of ionization ($x < 10^{-13}$), as has been pointed by many previous studies (e.g., Gammie, 1996; Sano et al., 2000). In other words, a quiet area without turbulence appears around the midplane. In such a quiet area, solid particles can grow with low-velocity collisions ($< 10\,\mathrm{m\,s^{-1}}$) . Furthermore, recent numerical simulations suggest that the turbulence completely disappears even near the surface, where the degree of ionization is high, if taken into account both the ambipolar diffusion and the Hall effect (Simon et al., 2013a,b; Bai and Stone, 2013a,b, 2014; Bai, 2013, 2015; Lesur et al., 2014; Kunz and Lesur, 2013).

In addition, in the case of **Assumption B**, Okuzumi et al. (2012) found that low-velocity collisions form porous aggregations with low density ($\rho < 10^{-2} - 10^{-5}\,\mathrm{g\,cm^{-3}}$), instead of compact particles with $\sim 1\,\mathrm{g\,cm^{-3}}$, if the impact energy is lower than the threshold energy, $E_{\mathrm{roll}}$, determined by material properties (Blum and Wurm, 2000; Suyama et al., 2008; Okuzumi et al., 2009). Suyama et al. (2008) found that the density of such a porous aggregation does not change even if the collision has higher energy than $E_{\rm roll}$ based on numerical simulations. Because of the relatively low density of the porous aggregations, their in-spiral velocity can be low enough to stay in the disk for a long time.

Yet in the case of **Assumption C**, several authors pointed out that a real accretion disk contains density discontinuities at the water sublimation zone (Kretke and Lin, 2007) and at the outer edge of an MRI inactive zone (Daisaka et al., 2006; Masset et al., 2006; Matsumura et al., 2007). At these locations, the torques cause outward type-I migration, which generate traps for migrating embryos.

Furthermore, Ormel and Klahr (2010); Lambrechts and Johansen (2012) proposed that the accretion of pebbles with size of 1-100 cm can be high enough to corroborate **Assumption D**, since they rapidly loose their excess energy and angular momentum through the interaction with gas. Indeed, substantial amounts of solid particles are expected to be mostly in the form of pebbles (Lambrechts and Johansen, 2012). Although it is likely to help the formation of gas giants and icy planets (Neptune and Uranus; Lambrechts and Johansen (2014)), the formation of terrestrial planets is unclear. In fact, according to the simulation of planet formation by Kretke and Levison (2014), hundreds of Mars- and Earth-mass objects between 4-10 AU are formed, instead of several giant planet cores larger than 10 Earth-mass, which are necessary to capture gas to become a gas giant. Such a system is not likely to be a planetary system similar to our solar system. Although Levison et al. (2015) found that this difficulty would overcome if pebbles formation continued in the disk during $10^7$ yrs, though any reasons of the continued formation of pebbles are not known.

Hansen (2009) assumed a narrow annulus composed of 400 planetesimals between 0.7 and 1.0 AU for the planetesimal-growth stage to construct a planet system similar to the solar system. The total mass of the planetesimals in the torus is set to be 2 Earth masses, and the corresponding average-column density is about 10 times higher than that of MMSN model. They found that four terrestrial planets are produced just like our solar system in most cases. In order to explain this rather artificial distribution of planetesimals, Walsh et al. (2011) and Walsh and Morbidelli (2011) used outer gas planets, Jupiter and Saturn. They assumed that Jupiter first forms with subsequent migration inward down to 1.5-2 AU. Then, Saturn grows and migrates to be eventually locked in 2:3 resonance with Jupiter. The resonated pair of the planets migrates outward back to 5 and 7 AU. Jupiter directs planetesimals inward by gravitational scattering to form a narrow torus of planetesimals around 1 AU. This model, named the "Grand Tack Model", has become popular, though it is theoretically uncertain how Jupiter can be made in the outer solar system, where planetesimal growth is slow. Furthermore, "Grand Tack Model" has a difficulty to clean-up the inner region than 0.7 AU, where the solid component also shows a significant deficit.

In addition, many exoplanets (2041 planets as of 6 January 2015; http://exoplanet.eu/catalog/) have been discovered through observational techniques (e.g., Udry and Santos (2007)), revealing that planetary systems have much greater variety than previously thought before their discoveries. Ida and Lin (2004a,b, 2005, 2007); Ida and Lin (2008) have performed systematic study with the standard model through a series of population-synthesis simulations, finding that the planetary mass ($M_{\rm p}$)-semi-major axis (a) distribution observed from exoplanetary systems can be reproduced, if "type I migration reduction factor", $C_1$, would be as small as $0.03 - 0.1$. Unfortunately, the real number of $C_1$ is not known yet, though many different formulas have been produced for various situations (Paardekooper and Papaloizou, 2008, 2009a,b; Paardekooper et al., 2010, 2011; Paardekooper, 2014).

In this work, we have constructed a steady-state, 1-D model of accretion disk around a newly born star, based on the $\alpha$-model of accretion disk (Shakura and Sunyaev, 1973) with a given accretion rate $\dot{M}$ ranging from $10^{-6.5}$–$10^{-8.0}\,{\rm M}_\odot\,{\rm yr}^{-1}$. In young stellar objects, $\dot{M}$ decreases with a timescale of $\sim 10^6$ yrs. It allows us to explore a wider parameter of space compared with the models restricted to MMSN. In addition, we take into account of magneto-rotational instability (Balbus and Hawley, 1991; Hawley and Balbus, 1991) and porous aggregation of solid particles (Okuzumi et al., 2012; Kataoka et al., 2013) as well as the ionization of thermal excitation inside of the disk. The thermal ionization of the alkali metal atoms such as potassium (K) and sodium (Na) become important when the temperature raises above 1000 K (Pneuman and Mitchell, 1965; Umebayashi and Nakano, 1981, 1988). The cosmic rays penetrate deep inside the gas disk, contribute to the ionization of the gas along with the ionizing radiation of radioactive nuclei in the gas (Umebayashi and Nakano, 2009).

We found that the disk consists of three regions: Outer Turbulent Region (OTR: $r > r_{\rm out}$), MRI suppressed region (MSR: $r_{\rm in} < r < r_{\rm out}$), and Inner Turbulent Region (ITR: $r_{\rm A} < r < r_{\rm in}$; see figure 1). The OTR outside of

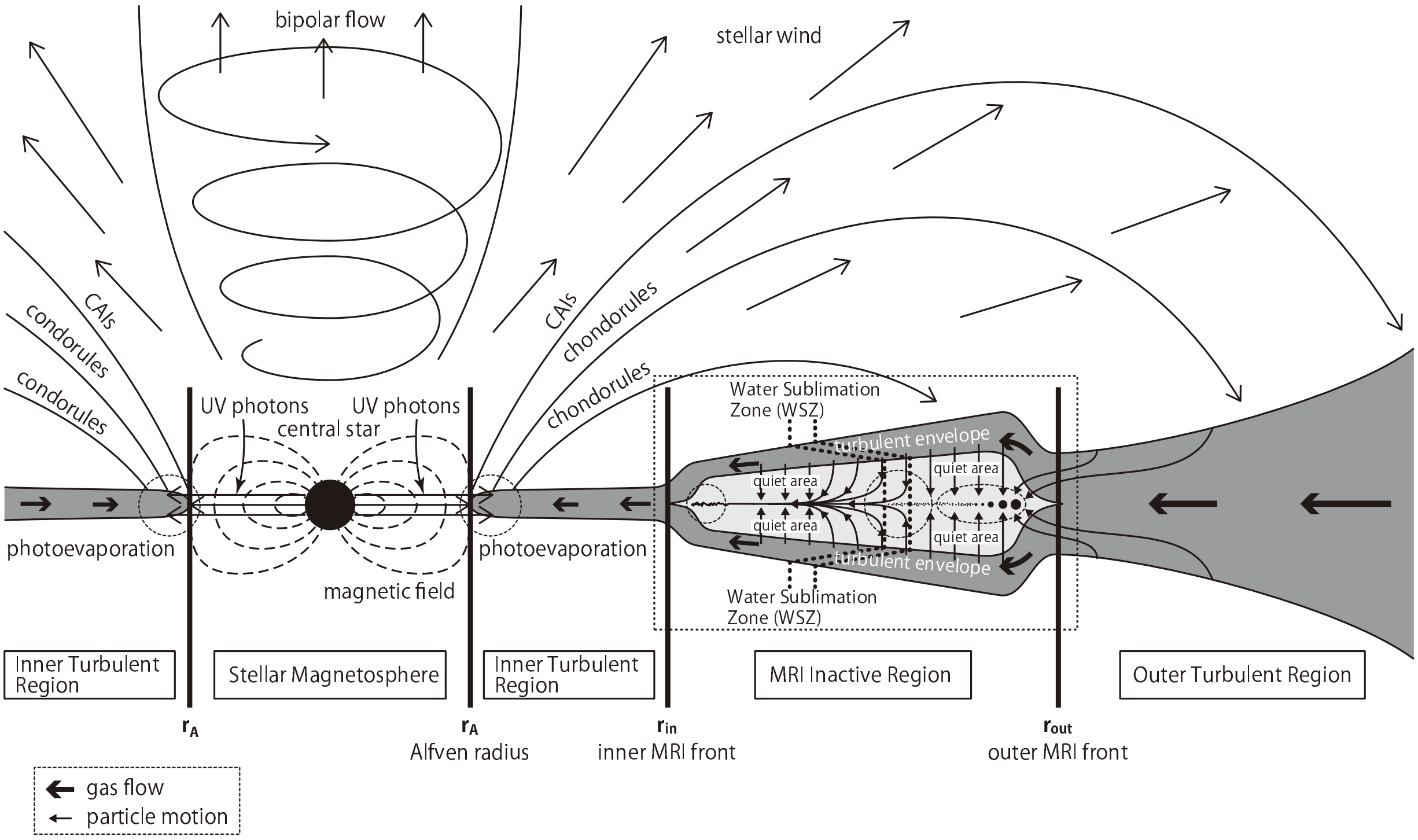

Figure 1: Schematic cross section of the structure of the protoplanetary disk proposed in the present paper. See text for detail.

$r_{\rm out}$ ($\sim 8-60\,$AU) is fully turbulent due to Magneto-Rotational Instability, which is generally believed to be a major source of turbulent viscosity in the accretion disk (Sano et al., 2000). However, in $r < r_{\rm out}$, a quiet area without turbulence appears around the midplane of the gas disk, since the ionization degree is so low that MRI is suppressed. It is sandwiched by turbulent envelopes, where the degree of ionization remains high enough due to the comic rays. The disk becomes fully turbulent (Balbus and Hawley, 2000) again around $r = r_{\rm in}$ ($\sim 0.2-1$AU), and the inner turbulent region initiates, because the midplane temperature becomes high enough ($> 1000$ K) through gravitational energy release to ionize alkali (K and Na) atoms (Pneuman and Mitchell, 1965; Umebayashi and Nakano, 1981, 1988). The inner turbulent region continues to the Alfven radius ($r_{\rm A} = 0.01-0.03$AU), where the stellar magnetic field is strong enough to truncate the gas disk. Such a three-region structure is consistent with the view of the recent comprehensive review by Armitage (2011). The quiet area (QA) is the most interesting in the context of planetesimal formation (figure 2). We found that planetesimals are formed at two distinct areas around the outer ($r \sim r_{\rm out}$) and inner MRI fronts ($r \sim r_{\rm in}$), as we will explain later in the present paper in detail. The gas giants and ice planets are likely to be made in the former, while the terrestrial planets in the latter. First, porous icy aggregations can grow as shown by Okuzumi et al. (2012). We found that they reach $10^{27}$g $\sim \mathrm{M}_\oplus$ and beyond within a million years at the outer MRI front.

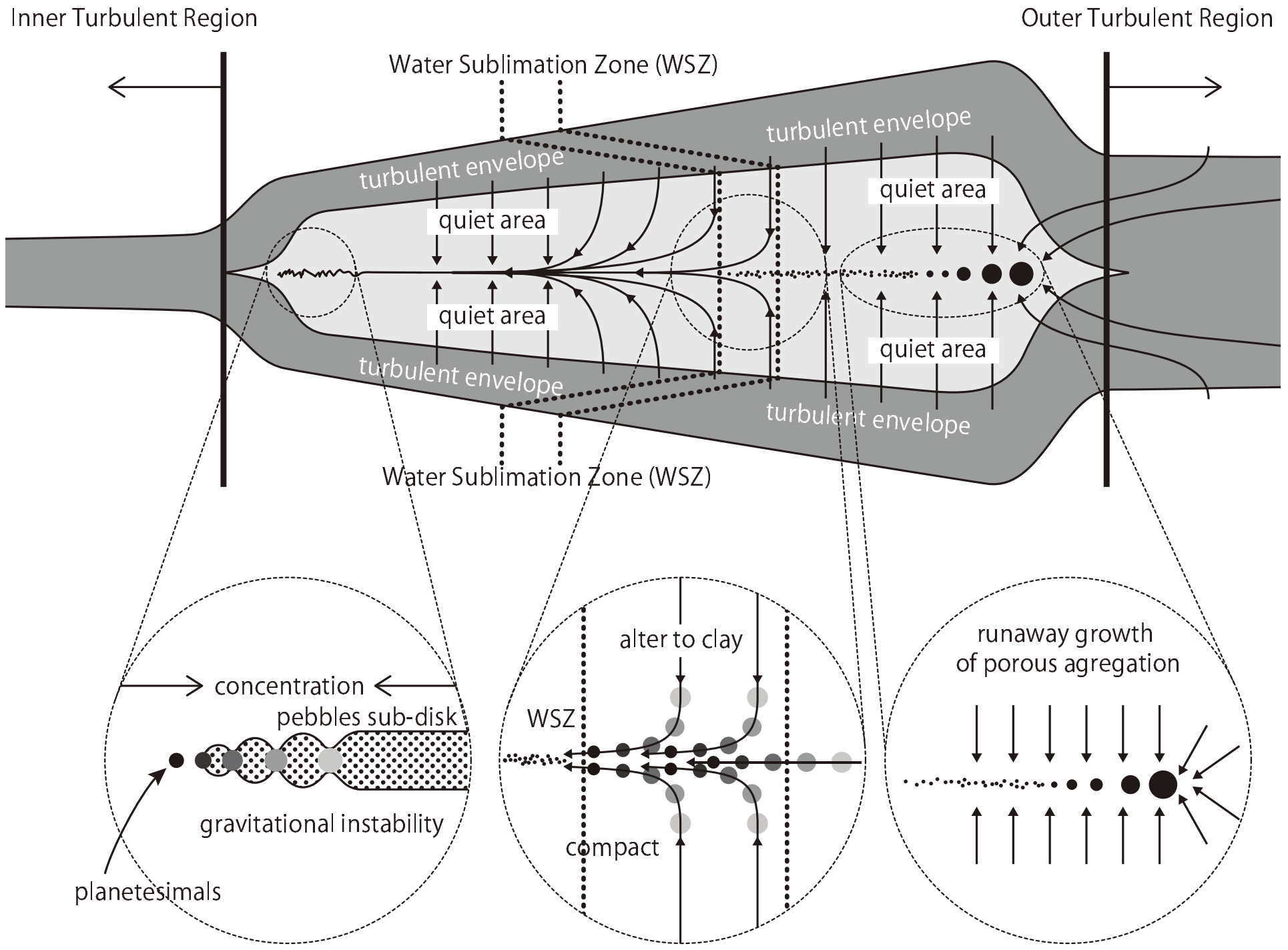

Figure 2: At the outer and inner boundaries of the quiet area (the outer and inner MRI fronts), solid particles are accumulated to form icy and rocky planetesimals, respectively. In the water sublimation zone (WSZ), icy porous aggregations are altered and compacted to dense clay minerals.

On the other hand, rocky planetesimals are formed around the inner MRI front, since the drift velocity of pebbles changes the direction from inward to outward, because of the positive pressure gradient $\partial P/\partial r$ respect to $r$. Pebbles are trapped and accumulated around $r_{\rm in}$, and undergo gravitational instability to form km-sized planetesimals (Johansen et al., 2006; Johansen et al., 2007; Johansen and Youdin, 2007; Johansen et al., 2009; Johansen and Lacerda, 2010; Johansen et al., 2011; Lambrechts and Johansen, 2014; Chatterjee and Tan, 2014, 2015). These planetesimals can grow further through the accretion of pebbles (Ormel and Klahr, 2010; Lambrechts and Johansen, 2012, 2014). The pebbles and planetesimals can remain there for a long time, since they are trapped by outward torques due to a positive density gradient, which always exists there (Kato et al., 2009, 2010), while gas steadily spirals inward towards the central star.

In summary, we have successfully constructed a qualitative model of planetesimal formation overcoming difficulties presented above. We name it the tandem planet formation regime, since there are two distinct sights of the planetary formation in the disk.

The tandem planet formation regime is important to form a planet fostering life, like the Earth. First, Maruyama and Ebisuzaki (2016, private communication) suggested that the Earth was born as a naked planet without water/volatiles. Interestingly enough, the plenetesimals formed in the inner MRI front are completely volatile free in the tandem regime, since the site of the inner planetesimal formation is as hot as 1000-1300 K and physically separated from those of ice-rich planetesimals. On the other hand, in the previous models, planetesimals are formed continuously in the entire region (0.5-20 AU) of the gas disk, so that the mixing of the materials is inevitable. Only the tandem planet formation is, therefore, consistent with the view that Earth was born as a naked planet. Second, Ebisuzaki and Maruyama (2016, private communication) noted that rich uranium ores can drive a natural nuclear reactor to produce organic materials after water was seeded on the Earth through the late heavy bombardment of volatile-rich asteroids. Furthermore, the phosphate-bearing minerals produced through the reductive and completely dry conditions of the Hadean Earth (such as schreibersite: $Fe_3P$) can react actively with water to supply phosphate in a biologically-active form. Finally, the tandem planet formation is consistent with the concept of Habitable Trinity (Dohm and Maruyama, 2015), in which the amount of water at the surface of the Earth must be at a sufficient level to foster life; three essentials (Habitable Trinity) for life, i.e. atmosphere, ocean, and landmass interacting due to Sun-driven hydrological cycling.

The present paper is the first paper of a series that describe a new overarching framework of planetary formation. Hear we explain tandem planet formation regime in which a solar system like planetary systems are likely formed. In the following, we begin by describing the assumptions of our model (Section 2). This is followed by details of the structure and evolution of the protoplanetary disk based on the results of a 1-D accretion model (Section 3) and a planetesimal-formation mechanism (Section 4). Our results are then compared with those of other models and observations (Section 5).

## 2. Model and Assumptions

Protoplanetary disks are geometrically thin, with the disk thickness being much smaller than the radius $r$ from the central star. In such a case, the disk can be approximated by an 1-D model, characterized by z-integrated value, such as $\Sigma$ ($= 2\int_0^\infty \rho dz$) and midplane temperature (and density). Here, the vertical-density profile (z-) is assumed to be the Gaussian function with the scale height of $H$, i.e., $\rho \propto \exp(-(z/2H)^2)$.

### *2.1. Steady State Model of 1D gas disk*

The time evolution of the column density $\Sigma$ is described by:

$$\frac{\partial \Sigma}{\partial t} = \frac{1}{2\pi r}\frac{\partial \dot{M}}{\partial r}, \tag{1}$$

where $\dot{M}$ is the mass accretion rate (positive for inward accretion). For the steady-state case, i.e., $\partial/\partial t = 0$, we obtain

$$\frac{\partial \dot{M}}{\partial r} = 0. \tag{2}$$

The mass accretion rate is expressed as:

$$\dot{M} = -2\pi r \Sigma v_r = 6\pi r^{1/2}\frac{\partial}{\partial r}(\Sigma \nu r^{1/2}) = \text{const.}, \tag{3}$$

where $v_r$ is the radial velocity of the gas in the disk, and $\nu$ is the disc viscosity. In the $\alpha$ disk assumption (Shakura and Sunyaev, 1973), the viscosity is given as:

$$\nu = \bar{\alpha} c_{\rm s} H, \tag{4}$$

where $H = c_{\rm s}/\Omega$ is the vertical scale height of the disk, $c_{\rm s}$ is the isothermal sound velocity, and $\Omega$ is the Keplarian orbital frequency. These are given by:

$$H = c_{\rm s}/\Omega = \left(\frac{k_{\rm B} T_{\rm m} r^3}{\mu m_H G M_*}\right)^{1/2} = 5.0\times 10^{11}\left(\frac{T_{\rm m}}{280\,{\rm K}}\right)^{1/2}\left(\frac{r}{\rm AU}\right)^{3/2} \quad {\rm cm} \tag{5}$$

$$c_{\rm s} = \left(\frac{k_{\rm B} T_{\rm m}}{\mu m_{\rm H}}\right)^{1/2} = 9.9\times 10^{4}\left(\frac{T_{\rm m}}{280\,{\rm K}}\right)^{1/2} \quad {\rm cm\,s^{-1}} \tag{6}$$

$$\Omega = \left(\frac{GM_*}{r^3}\right)^{1/2} = 1.99 \times 10^{-7} \left(\frac{r}{\mathrm{AU}}\right)^{-3/2} \quad \mathrm{s}^{-1}, \tag{7}$$

where $T_{\mathrm{m}}$ is the midplane temperature of the disk, $k_{\mathrm{B}}$ is the Boltzmann constant, $\mu = 2.34$ is the mean molecular weight of gas, $m_{\mathrm{H}}$ is the mass of an hydrogen atom, $G$ is the gravitational constant, and $M_*$ is the mass of the central star. In the calculation, the gas disk is represented by 8000 cells equally divided from the stellar surface ($R_* = 3R_\odot = 0.014$ AU) to 100AU in log radial space.

*2.2. Midplane Temperature*

Taking into account of the gravitational energy release, the midplane temperature $T_{\mathrm{m}}$ is given by (Hubeny, 1990; Kretke and Lin, 2007):

$$T_{\mathrm{m}}^4 = \left(\frac{3\dot{M}\Omega^2}{8\pi\sigma}\right)\left(\frac{3}{8}\tau + \frac{\sqrt{3}}{4}\right) + T_{\mathrm{irr}}^4, \tag{8}$$

where $\sigma$ and $\tau$ are respectively the Stefan-Boltzmann constant and the optical depth at the midplane (see subsection 2.4). Here, $T_{\mathrm{irr}}$ is the irradiation temperature given by:

$$T_{\mathrm{irr}}^4 = \frac{1}{2}\left(1-\epsilon\right) T_*^4 \left(\frac{R_*}{r}\right)^2 \left[\frac{4}{3\pi}\left(\frac{R_*}{r}\right) + \frac{2}{7}\frac{H}{r}\right], \tag{9}$$

where $\epsilon = 0.5$ is the disk albedo (Coleman and Nelson, 2014), $T_* = 4000\,\mathrm{K}$ is the stellar temperature, and $R_* = 3R_\odot$ is the stellar radius, respectively. The equation 8 is solved by iteration.

*2.3. Onset Condition of Magneto-Rotational Instability (MRI)*

The magneto-rotational instability takes place when Elsasser number $\Lambda$ is larger than unity (Jin, 1996; Sano and Miyama, 1999; Turner et al., 2007). It is defined as:

$$\Lambda = \frac{V_{\mathrm{A,z}}^2}{\eta_{\mathrm{O}}\Omega} = \frac{2}{\beta_{\mathrm{Z}}}\frac{c_{\mathrm{s}}}{\eta_{\mathrm{O}}}H = 2.6 \times 10^{-1}\left(\frac{x}{10^{-12}}\right)\left(\frac{\beta_{\mathrm{Z}}}{100}\right)^{-1}\left(\frac{T_{\mathrm{m}}}{280\,\mathrm{K}}\right)^{1/2}\left(\frac{r}{\mathrm{AU}}\right)^{3/2}, \tag{10}$$

where $\eta_{\mathrm{O}} = 2.34 \times 10^3 (T/100\mathrm{K})^{1/2} x^{-1} \mathrm{cm}^2\,\mathrm{s}^{-1}$ is the Ohmic resistivity (Sano et al., 2000), $V_{\mathrm{A,z}}$ is the z-component of Alfven wave velocity, and $\beta_{\mathrm{Z}} = 2c_{\mathrm{s}}^2/V_{\mathrm{A,z}}^2$. We determine z-averaged $\alpha$-value, $\bar{\alpha}$, depending on $\Lambda$ as:

$$\bar{\alpha} = \begin{cases} \alpha_{\rm act} & \text{for} \quad \Lambda > 1 \quad \text{(MRI active)} \\ \alpha_{\rm inact} & \text{for} \quad \Lambda < 1 \quad \text{(MRI inactive)}, \end{cases} \tag{11}$$

where we adopted $\alpha_{\rm act}$ as $1.0 \times 10^{-2}$, according as the numerical MHD simulations of planetary disks (Davis et al., 2010; Shi et al., 2010). As for $\alpha_{\rm inact}$, we use $\gamma\alpha_{\rm act}$. Here $\gamma$ is the reduction factor of turbulance due to the suppression of MRI, taking into account of the quiet area without turbulence around the midplane. In the MRI inactive region, the disk is composed of a quiet area and two turbulent envelopes. The former is turbulent free, while the latter are fully turbulent, referred to as "layered accretion" (e.g., Gammie, 1996; Sano et al., 2000). Here, we assume $\gamma = 10^{-0.5} \simeq 0.316$ for the entire MRI inactive region, though $\gamma$ has not yet been constrained well. In fact, the recent numerical simulations suggest that the turbulence in MRI inactive regions completely disappear if we take into account the ambipolar diffusion and the Hall effect (Simon et al., 2013a,b; Bai and Stone, 2013a,b, 2014; Bai, 2013, 2015; Lesur et al., 2014; Kunz and Lesur, 2013). In such a case, $\gamma$ can be as small as $10^{-5}$. Even in such a case, the choice of $\gamma$ does not change the onset condition of MRI, which is mainly determined by $\alpha_{\rm act}$, though $\alpha_{\rm inact}$ determines the column density in the MRI inactive region.

The ionization degree $x$ is evaluated as:

$$x = \max(x_{\rm th}, x_{\rm i}), \tag{12}$$

where $x_{\rm th}$ is the ionization degree determined by thermal equilibrium, and $x_{\rm i}$ is that determined by the ionization equilibrium. First, $x_{\rm th}$ is calculated as:

$$x_{\rm th} = n_{\rm n}^{-1/2} \left( \frac{2\pi m_{\rm e} k_{\rm B} T_{\rm m}}{h^2} \right)^{3/4} \left[ f_{\rm K}^{1/2} \exp\left( -\frac{E_{\rm K}}{2k_{\rm B}T_{\rm m}} \right) + f_{\rm Na}^{1/2} \exp\left( -\frac{E_{\rm Na}}{2k_{\rm B}T_{\rm m}} \right) \right], \tag{13}$$

for the case that the density and the temperature of the gas are sufficiently high, where local thermal equilibrium condition is satisfied, and the ionization degree is determined by the Saha equation (e.g., Balbus and Hawley, 2000). Here, $h$ is the Plank constant, and $n_{\rm n}$ is the number density of neutral molecules,

$$n_{\rm n} = \frac{\Sigma}{\sqrt{2\pi} H \mu m_{\rm H}}. \tag{14}$$

We adopted the following ionization energy and the abundance of alkali metals:

$$\begin{array}{lclclcl} E_{\rm K} & = & 4.341 \text{ eV}, & & f_{\rm K} & = & 1.35 \times 10^{-7}, \\ E_{\rm Na} & = & 5.139 \text{ eV}, & & f_{\rm Na} & = & 2.06 \times 10^{-6}. \end{array} \quad (15)$$

On the other hand, $x_{\rm i}$ is calculated by (Okuzumi, 2009):

$$x_{\rm i} = \frac{\alpha_{\rm p} n_{\rm p}}{2\alpha_{\rm g} n_{\rm n}} \left( \sqrt{1 + \frac{4\alpha_{\rm g} \mu n_{\rm n}}{\alpha_{\rm p}^2 n_{\rm p}^2} \zeta} - 1 \right), \quad (16)$$

assuming ionization equilibrium, in which the ionization rate $\zeta$ and the recombination rate $\chi$ are balanced (Pneuman and Mitchell, 1965; Umebayashi and Nakano, 1988; Umebayashi and Nakano, 2009). Here, $\alpha_{\rm p}$ and $\alpha_{\rm g}$ are the recombination coefficients respectively in solid particles and in gas, which will be given later. The number density $n_{\rm p}$ of solid particles is given by:

$$n_{\rm p} = \frac{3\Sigma \bar{f}_{\rm p}}{\sqrt{32\pi^3} H \bar{a}_{\rm p}^3 \bar{\rho}_{\rm i}}, \quad (17)$$

where $\bar{a}_{\rm p} = 10\,\mu$m is the radius of particle, and $\bar{f}_{\rm p}$ and $\bar{\rho}_{\rm i}$ are respectively the particle fraction in mass and the internal density of the particles. Here, we assume that $\bar{\rho}_{\rm i} = 1.0$ g cm$^{-3}$ for $T < 150$ K, 2.0 g cm$^{-3}$ for $180 < T < 1380$ K, and 0 g cm$^{-3}$ for $T > 1380$ K, and $\bar{f}_{\rm p} = 1.0\times10^{-2}$ for $T < 150$ K, $2.5\times10^{-3}$ for $180 < T < 1380$ K, and 0 for $T > 1380$ K (Hayashi, 1981; Hayashi et al., 1985). They are linearly interpolated in the water sublimitation zone ($150 < T < 180$ K).

Here, we do not take into account the particle growth in the structure of the gas disk for simplicity. It is justified by the fact that the BCCA growth of porous aggregation does not change the properties of the solid particles in the respect to recombination and that the disk always comprises a turbulent area in which particle growth is likely to be insignificant.

According to Umebayashi and Nakano (2009), the ionization rate (galactic cosmic rays, radionuclides, and thermal ionization) at the midplane of the disk is calculated as:

$$\zeta = 2\zeta_{\rm CR}(0) \left[ 1 + \left( \frac{\Sigma}{2\Sigma_{\rm p,10GeV}} \right)^{3/4} \right]^{-4/3} \exp\left( -\frac{\Sigma}{2\Sigma_{\rm p,10GeV}} \right) + \zeta_{\rm RN} + \zeta_{\rm th}, \quad (18)$$

where $\Sigma_{\rm p,10GeV} = 96\,{\rm g\,cm^{-2}}$ is the range of the 10 GeV protons for hydrogen gas, $\zeta_{\rm CR}$ is the ionization rate due to galactic cosmic rays ($1.0 \times 10^{-17}\,{\rm s^{-1}}$), and $\zeta_{\rm RN}$ is the ionization rate due to radionuclides ($7.6 \times 10^{-19}\,{\rm s}^{-1}$ for short-lived nuclides and $1.4 \times 10^{-22}\,{\rm s}^{-1}$ for long-lived nuclides). When the midplane temperature is higher than 1000 K, the thermal ionization process becomes important. The thermal ionization rate, due to potassium and sodium, is given by $\zeta_{\rm th} = \zeta_{\rm th,K} + \zeta_{\rm th,Na}$. These rates are estimated as:

$$\zeta_{\rm th,Na} = \pi a_{\rm Na}^2 f_{\rm Na} n_{\rm n} \left(\frac{8k_{\rm B}T_{\rm m}}{\pi\mu m_{\rm H}}\right)^{1/2} \exp\left(-\frac{E_{\rm Na}}{k_{\rm B}T_{\rm m}}\right) \tag{19}$$

$$\zeta_{\rm th,K} = \pi a_{\rm K}^2 f_{\rm K} n_{\rm n} \left(\frac{8k_{\rm B}T_{\rm m}}{\pi\mu m_{\rm H}}\right)^{1/2} \exp\left(-\frac{E_{\rm K}}{k_{\rm B}T_{\rm m}}\right), \tag{20}$$

where we use

$$a_{\rm Na} = 0.116\ {\rm nm}, \qquad \text{and} \qquad a_{\rm K} = 0.152\ {\rm nm}. \tag{21}$$

The rate coefficient $\alpha_{\rm p}$ of the recombination on the surface of a particle is calculated as:

$$\alpha_{\rm p} = (1-y)\pi\bar{a}_{\rm p}^2 v_{\rm i}. \tag{22}$$

Here, $v_{\rm i} = (8k_{\rm B}T_{\rm m}/\pi m_{\rm i})^{1/2}$ is the thermal velocities of ions, and the coefficient $y$ can be obtained when the charge $Z$ of the solid particle satisfies the ordinary differential equation (Okuzumi, 2009):

$$\frac{dZ}{dt} = \pi\bar{a}_{\rm p}^2 \left[v_{\rm i}(1-\frac{q^2Z}{\bar{a}_{\rm p}k_{\rm B}T_{\rm m}}) - v_{\rm e}\exp(\frac{q^2Z}{\bar{a}_{\rm p}k_{\rm B}T_{\rm m}})\right], \tag{23}$$

where $v_{\rm e} = (8k_{\rm B}T_{\rm m}/\pi m_{\rm e})^{1/2}$ is the thermal velocities of electrons. The charge equilibrium of the solid particles is established when:

$$v_{\rm i}(1-\frac{q^2Z}{\bar{a}_{\rm p}k_{\rm B}T_{\rm m}}) - v_{\rm e}\exp(\frac{q^2Z}{\bar{a}_{\rm p}k_{\rm B}T_{\rm m}}) = 0. \tag{24}$$

Then the equation of $(v_{\rm i}/v_{\rm e})(1-y) = e^y$ has a solution of $y = \frac{q^2Z}{\bar{a}_{\rm p}k_{\rm B}T_{\rm m}} \sim -2.8$, when $v_{\rm i}/v_{\rm e} = \sqrt{m_{\rm e}/m_{\rm i}} = 1.65 \times 10^{-2}$.

The recombination coefficient $\alpha_{\rm g}$ in gas can be approximated by that of dissociative recombination (McCall et al., 2004):

$$\alpha_{\rm g} = 6.7 \times 10^{-8} \left(\frac{T_{\rm m}}{300\,{\rm K}}\right)^{-1/2} {\rm cm}^3\,{\rm s}^{-1}. \tag{25}$$

*2.4. Optical Depth*

Optical depth $\tau$ is given by:

$$\tau = \begin{cases} \kappa(T_{\rm m})\Sigma/2 & \text{for} \quad \text{MRI active} \\ \kappa(T_{\rm m})\Sigma_{\rm TE}/2 & \text{for} \quad \text{MRI inactive,} \end{cases} \tag{26}$$

where $\Sigma_{\rm TE} = \gamma\Sigma$ is the column density of the turbulent envelopes in the MRI inactive region. We adopted the opacity formula of Stepinski (1998) as shown in figure 3.

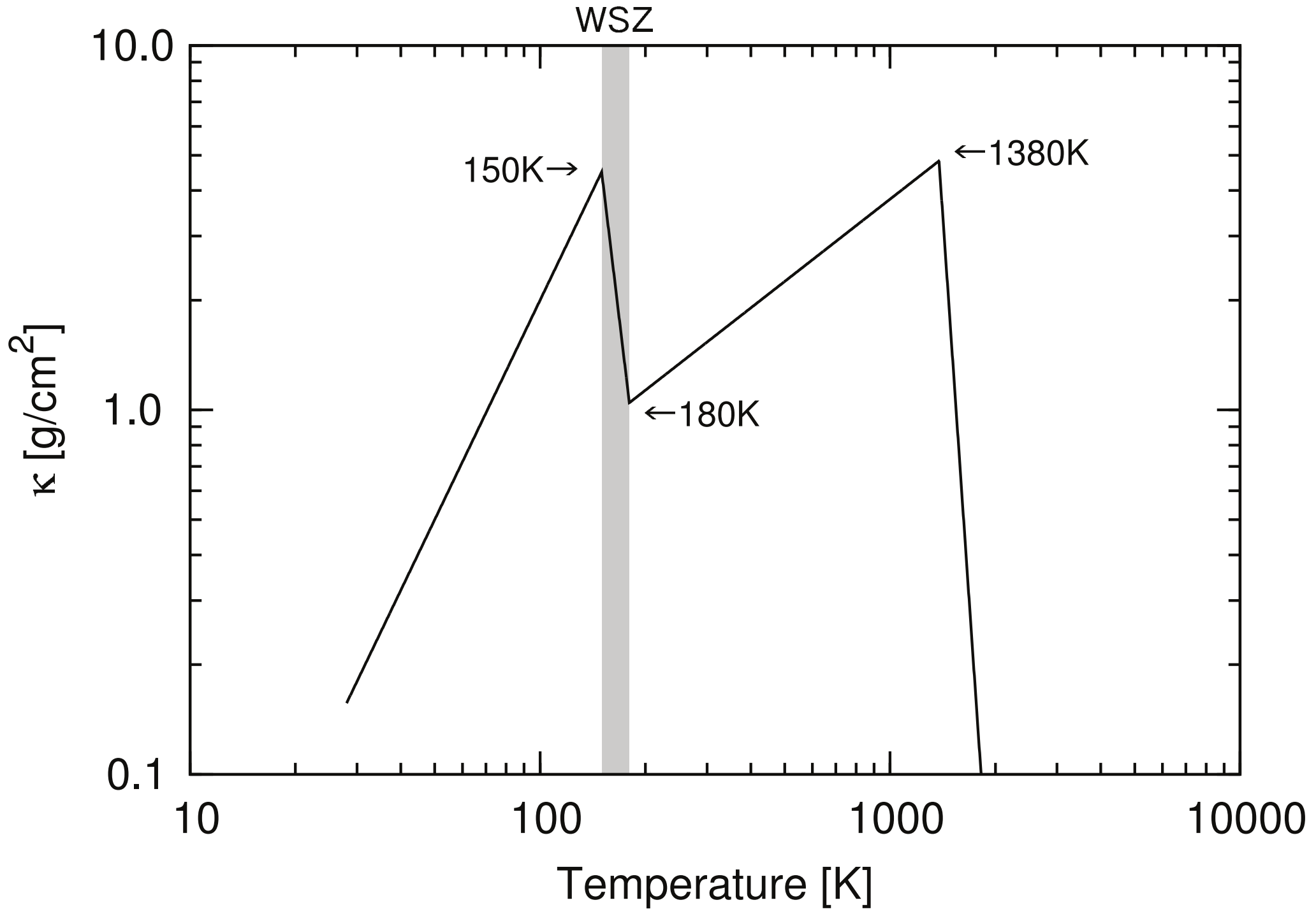

Figure 3: The opacity of the disk is adopted from Stepinski (1998). Hatched area represents water sublimation zone (WSZ).

*2.5. Vertical Magnetic Field*

For steady-state, vertical magnetic field $B_{\rm z}$ can be calculated as (Okuzumi et al., 2014):

$$B_{\rm z} = \frac{2\pi D}{c} K_\phi, \tag{27}$$

where the dimensionless coefficient $D$ is defined as:

$$D = -\frac{2}{\int_{-H}^{H} (v_{\rm r}/\eta_{\rm m}) dz}, \tag{28}$$

where $\eta_{\rm m}$ is the magnetic diffusivity (resistivity). Here, $K_\phi$ is the electric current integrated over the disk:

$$K_\phi = \int_{-H}^{H} J_\phi dz = \frac{c}{2\pi} B_{rs} - \frac{c}{4\pi} \frac{\partial}{\partial r} \int_{-H}^{H} B_z dz. \tag{29}$$

Here, $B_{rs} = B_r|_{z=H} (= -B_r|_{z=-H})$ is the radial field strength on the disk surface. The coefficient $D$ is evaluated as:

$$D \sim \frac{\eta_{\rm m}}{v_{\rm r} H}, \tag{30}$$

where $v_{\rm r} = \bar{\alpha} c_{\rm s}^2/(r\Omega)$ is the radial gas velocity. We assume that the turbulent (macroscopic) diffusivity $\eta_{\rm turb}$ dominates over the microscopic diffusivity $\eta_{\rm mol}$ in the entire region of the disk surface, in which case:

$$\eta_{\rm m} = \eta_{\rm turb}. \tag{31}$$

The turbulent diffusivity is obtained by (Lazarian and Vishniac, 1999):

$$\eta_{\rm turb} \sim v_{\rm turb} L_{\rm inj} \sim \sqrt{\bar{\alpha}} c_{\rm s} \xi \lambda_{\rm max}, \tag{32}$$

where $L_{\rm inj} = \xi \lambda_{\rm max}$ is the typical length of magnetic reconnection in the turbulence and $\lambda_{\rm max} = 2\pi v_{\rm A,z}/\Omega$ is the wavelength which gives the maximum growth rate of MRI (Balbus and Hawley, 1991).

Substituting equations 31 and 32 into equation 30, then $D$ is evaluated as:

$$D = \frac{8}{3} \pi \xi \left( \frac{2}{\bar{\alpha} \beta_{\rm z}} \right)^{1/2} \left( \frac{r}{H} \right). \tag{33}$$

Okuzumi et al. (2014) obtained the numerical solution of $B_{\rm z}$ distribution and found that $B_{\rm z}$ is approximated by the power law function of $r$, i.e. $r^{\nu_{\rm B}}$. The index $\nu_{\rm B}$ is about -2 for the case of $D \ll 1$ and $\nu_{\rm B} = 0$ for $D \gg 1$. Therefore, we may assume $\nu_{\rm B} = -2/(1 + D)$. In other words,

$$\frac{d \ln B_{\rm z}}{d \ln r} = -\frac{2}{1 + D}. \tag{34}$$

The value of $\xi$ depends on the nature of the magnetic turbulence, which is not well constrained, except that it takes a small value in well-developed turbulence. Here, we inwardly integrate equation 34 from $B_z = 3.4 \times 10^{-4}$ G at 100 AU, assuming $\xi \sim 3 \times 10^{-3} = const.$ It is consistent with the canonical value of the interstellar magnetic field ($\sim 10^{-6}$ G, e.g., Crutcher, 2012), if we taking into account the conservation of the magnetic flux in nebular contraction. The magnetic field reaches $10^2$ G at the inner most edge of the gas disk, which is also consistent of the photospheric magnetic field of T Tauri stars (Bouvier et al., 2007).

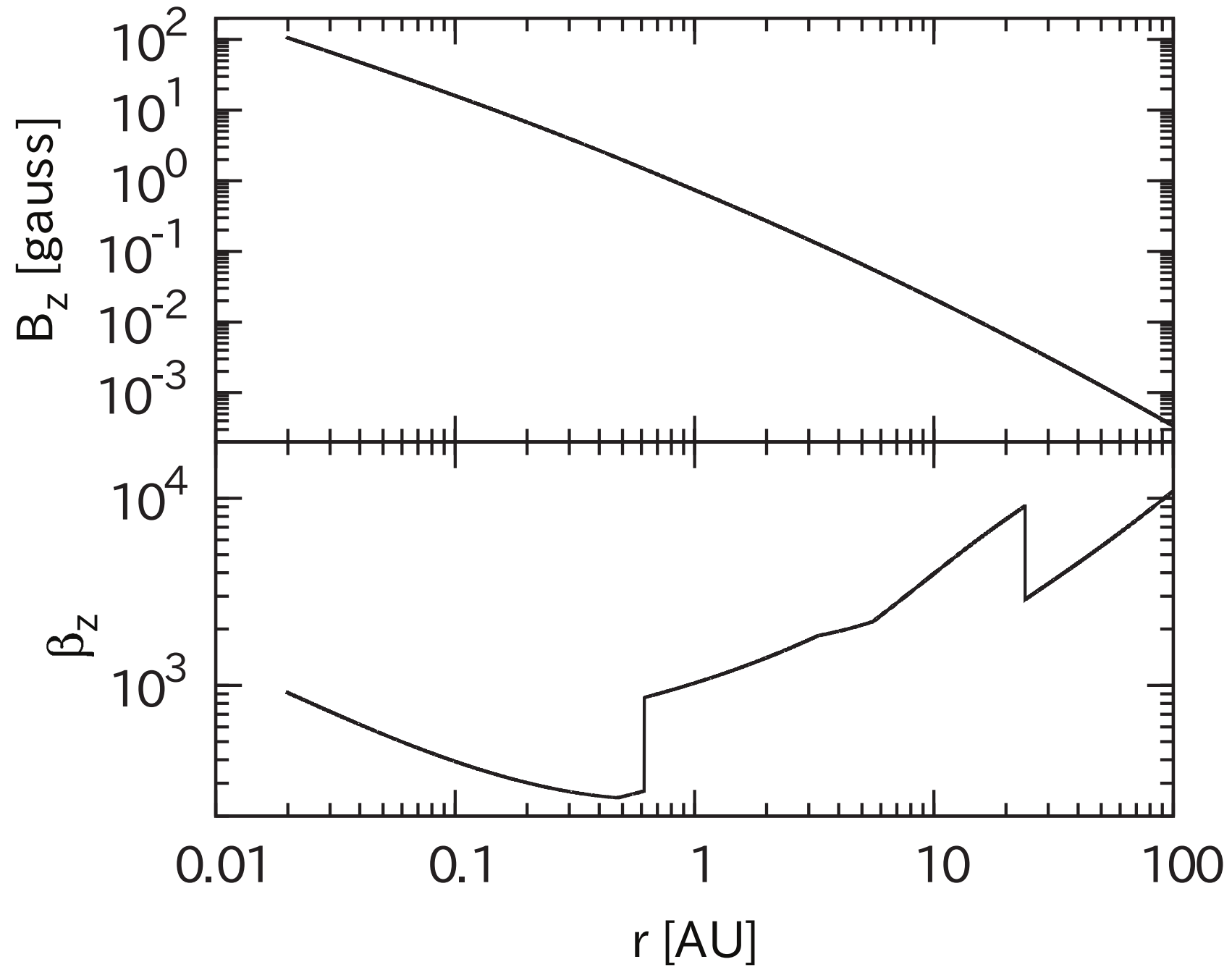

Figure 4: The vertical magnetic field $B_z$ and $\beta_z$ are plotted against the distance from the central star.

*2.6. Alfven Radius*

The gas disk is truncated by the magnetic field of the central star at the Alfven radius $r_A$, where magnetic pressure ($B^2/4\pi$) is equal to the rotational stress of the disk ($\rho_m v_K v_r$). It is calculated as:

$$r_A = \left( \frac{\mu_*^4}{G M_* \dot{M}^2} \right)^{1/7}, \tag{35}$$

where $\mu_*$ is the magnetic dipole moment of the central star (Shang et al., 2000).

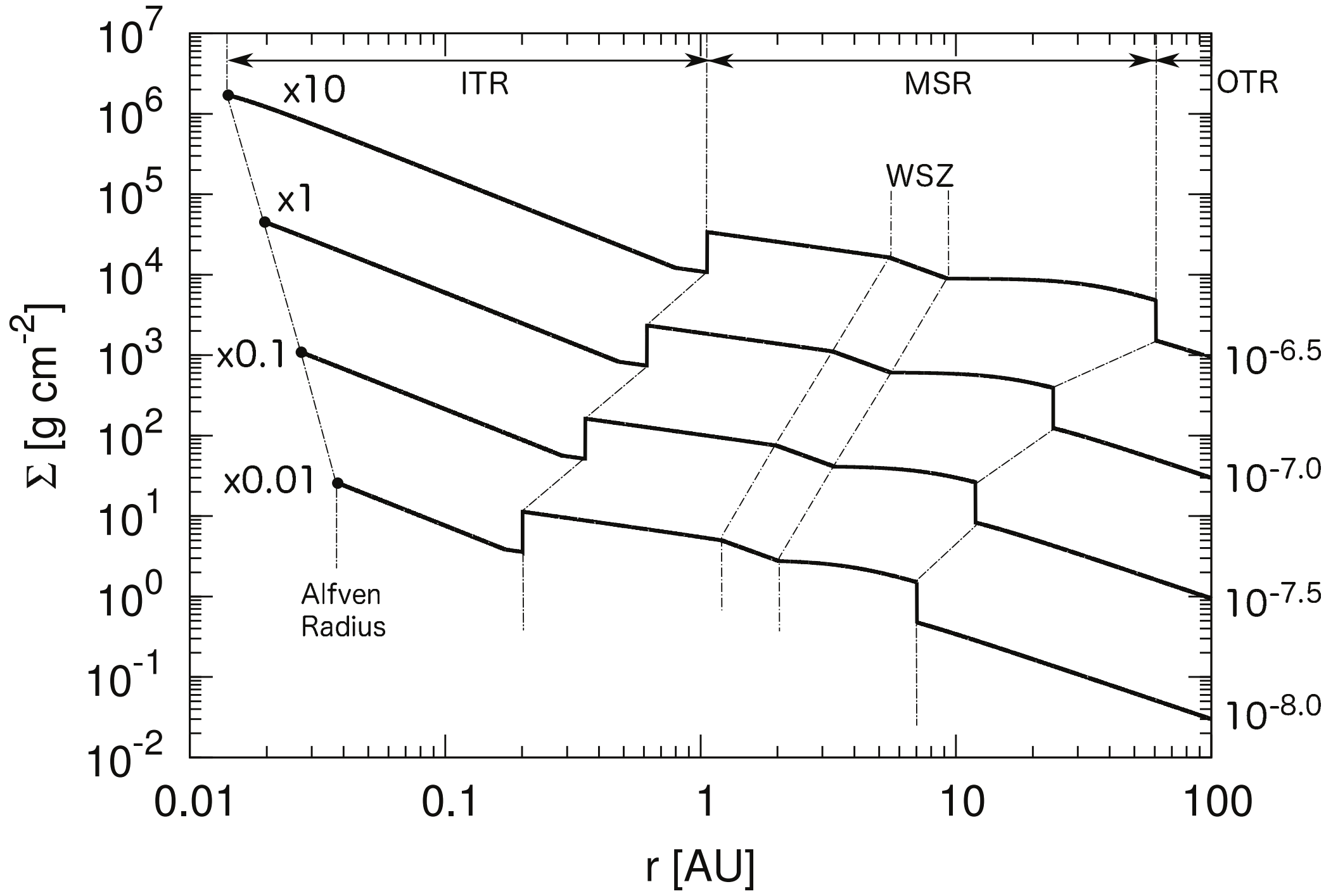

Figure 5: Radial profile of the column density $\Sigma$ of the gas in the disk for $\dot{M} = 10^{-6.5}, 10^{-7.0}, 10^{-7.5}$, and $10^{-8.0} \mathrm{M}_\odot\, \mathrm{yr}^{-1}$. A disk is divided into an outer turbulent region (OTR), an MRI suppressed region (MSR), and an inner turbulent region (ITR). In the MRI inactive region (MSR), the column density is $\sim \sqrt{\gamma}$ times higher than the turbulent regions (OTR and ITR) because of the $\sqrt{\gamma}$ times lower value of $\bar{\alpha}$. The black dots denote Alfven radii, where the disk is truncated by the stellar magnetic field.

## 3. Disk Structure

Figure 5 shows the radial profiles of the column density $\Sigma$ for different accretion rates ($\dot{M} = 10^{-6.5}, 10^{-7.0}$, $10^{-7.5}$, and $10^{-8.0}\, \mathrm{M}_\odot\, \mathrm{yr}^{-1}$). The disk can be divided into outer and inner turbulent regions (OTR and ITR), and an MRI suppressed region (MSR) (see also figures 1 and 5). As can be seen in figure 6, in the region $r_{\mathrm{in}} < r < r_{\mathrm{out}}$, both $\Lambda_{\mathrm{act}}$ and $\Lambda_{\mathrm{inact}}$ are less than unity, so that MRI is suppressed in the midplane of the disk. In figure 4, we plot vertical magnetic field $B_{\mathrm{z}}$ and $\beta_{\mathrm{z}}$ against distance from the central star.

### 3.1. Outer Turbulent region: OTR

Outer most region (Outer Turbulent Region: $r > r_{\rm out} = 8 \sim 60$ AU) is fully turbulent due to MRI: The ionization by cosmic rays and radioactive nuclei keep gas ionized enough to activate MRI.

### 3.2. MRI Suppressed Region: MSR

MRI is suppressed in the MRI suppressed Region (MSR), which is located between the outer and inner turbulent regions (OTR and ITR). MSR is vertically divided into three layers: the turbulent envelopes, the quiet area, and the sub-disk of pebbles. The surface is covered by two fully turbulent envelopes (figure 1), where the ionization degree is kept high enough to activate MRI. The gas flows inward through the turbulent envelopes (i.e., layered accretion: Gammie, 1996; Sano et al., 2000). A quiet area (QA) without turbulence appears near the midplane deep in the disk, where the ionization degree is low. The solid particles settle down to a relatively thin sub-disk of pebbles, as discussed later in section 4.

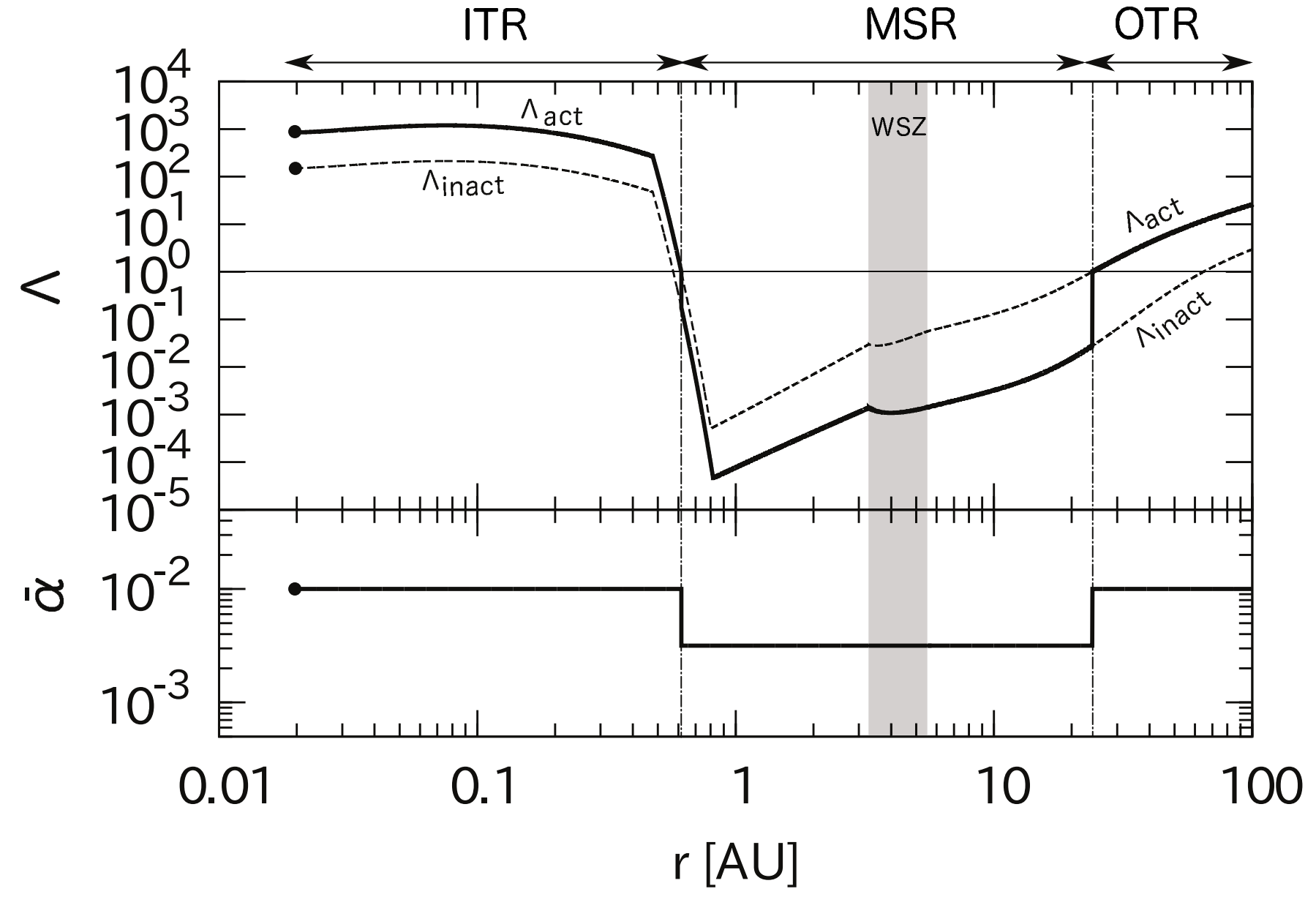

Figure 6: Radial profile of $\Lambda_{\rm act}$ (for MRI active) and $\Lambda_{\rm inact}$ (for MRI inactive) in the midplane of the steady-state solution of the protoplanetary disk with the accretion rate of $\dot{M} = 10^{-7.0}\,{\rm M}_\odot\,{\rm yr}^{-1}$. MRI is active if $\Lambda > 1$, otherwise inactive. We adopted a solid curve to construct the disk structures.

MRI suppressed region terminates when the midplane temperature raises higher than 1000 K: Thermal ionization becomes dominant rather than those of cosmic rays and radioactive nuclei. As a result, the Elsasser number increases rapidly, and once again reaches unity (figure 6).

*3.3. Inner Turbulent Region: ITR*

In addition to the outer turbulent region, there exists another fully turbulent region (inner turbulent region: $r < r_{\rm in} = 0.2 - 1$ AU), where the temperature of the midplane is high enough to activate MRI again because of thermal ionization of alkali metal atoms (Na and K). The disk is eventually truncated by the stellar magnetic field at Alfven radius ($r_{\rm A} = 0.01 - 0.03$ AU).

## 4. Tandem Planet Formation

We found that the outer and inner MRI fronts ($r_{\rm out}$ and $r_{\rm in}$) are the sites for planetesimal formation in the disk structure as described in the previous section (see also figure 2). In the present section, we discuss the growth of particles in the steady-state structure $\Sigma(r)$, $T_{\rm m}(r)$, $\rho_{\rm m}(r)$, $H(r)$, $c_{\rm s}(r)$, $\Omega(r)$, $H(r)$, and $\bar{f}_{\rm p}(r)$ of the disk with an accretion rate of $\dot{M} = 10^{-7.0}\,{\rm M}_{\odot}\,{\rm yr}^{-1}$.

*4.1. Particle Growth in a Tandem Disk*

*4.1.1. Lagrangian Superparticles*

In order to describe how the particles grow in the disk, we define a Lagrangian superparticle with three quantities ($r_{{\rm p},k}$, $z_{{\rm p},k}$, $m_{{\rm p},k}$). Here, $k$ is the superparticle index, $r_{{\rm p},k}$ the distance from the central star, $z_{{\rm p},k}$ the the scale height of the z-distribution of the particles, and $m_{{\rm p},k}$ the average mass of the particles that included in the superparticle. The total mass of particle associated to the superparticle $k$, is defined as $M_{{\rm p},k}(t=0) = \pi r_{{\rm s},k}(r_{{\rm s},k+1} - r_{{\rm s},k-1})\Sigma(r_{{\rm s},k})\bar{f}_{\rm p}(r_{{\rm s},k})$, where $r_{{\rm s},k} = r_{{\rm p},k}(t=0)$ is the starting radius of the superparticle.

The particle column density ($r = r_{{\rm p},k}$), the particle density at the midplane ($r = r_{{\rm p},k}, z = 0$), and the particle density at the particle scale height ($r = r_{{\rm p},k}, z = z_{{\rm p},k}$), are calculated by:

$$\Sigma_{{\rm p},k} = \frac{M_{{\rm p},k}}{2\pi r_{{\rm p},k}\Delta r_{{\rm p},k}}\frac{\bar{f}_{\rm p}(r_{{\rm p},k})}{\bar{f}_{\rm p}(r_{{\rm s},k})}, \tag{36}$$

$$\rho_{{\rm pm},k} = \frac{\Sigma_{{\rm p},k}}{\sqrt{2\pi}z_{{\rm p},k}}, \tag{37}$$

$$\rho_{{\rm p},k} = 0.6065\rho_{{\rm pm},k}. \tag{38}$$

The thickness of the superparticle in radius is

$$\Delta r_{\mathrm{p},k} = \frac{r_{\mathrm{p},k_+} - r_{\mathrm{p},k_-}}{2}, \quad (39)$$

where $r_{\mathrm{p},k_+}$ and $r_{\mathrm{p},k_-}$ are the outer and inner nearest particles of the particle $k$, respectively. The similar Lagrangian approach for the particle growth was proposed recently by (Krijt et al., 2015).

The three quantities are calculated by three ordinary differential equations:

$$\frac{dr_{\mathrm{p},k}}{dt} = -v_{r\mathrm{p},k},, \quad (40)$$

$$\frac{dz_{\mathrm{p},k}}{dt} = -v_{z\mathrm{p},k} \quad \text{for} \quad z_{\mathrm{p},k} > H_{\mathrm{pe},k}, \quad (41)$$

$$z_{\mathrm{p},k} = H_{\mathrm{pe},k} \quad \text{for} \quad z_{\mathrm{p},k} < H_{\mathrm{pe},k}, \quad (42)$$

$$\frac{dm_{\mathrm{p},k}}{dt} = \dot{m}_{\mathrm{p},k}, \quad (43)$$

where $v_{r\mathrm{p},k}$ is the radial particle velocity by hydro-dynamical interactions with gas, $v_{z\mathrm{p},k}$ is the particle settling velocity, and $H_{\mathrm{pe},k}$ is the particle scale height in equilibrium given by Youdin and Lithwick (2007):

$$H_{\mathrm{pe},k} = H(r_{\mathrm{p},k}) \left(1 + \frac{\Omega_k t_{\mathrm{s},k}}{\alpha_{\mathrm{D},k}}\right)^{-\frac{1}{2}} \left(1 + \frac{\Omega_k t_{\mathrm{s},k}}{\Omega_k t_{\mathrm{s},k} + 1}\right)^{-\frac{1}{2}}, \quad (44)$$

and $\Omega_k = \Omega(r_{\mathrm{p},k})$. The ordinary differential equations 43 are solved by the Euler method with the time increment $\Delta t$:

$$\Delta t = \min\left(0.05\frac{\Delta r_{\mathrm{p},k}}{|v_{r\mathrm{p},k}|}, 0.5\frac{m_{\mathrm{p},k}}{\dot{m}_{\mathrm{p},k}}\right). \quad (45)$$

The particle motion in the above gas disk is followed by 25 superparticles, which are distributed in logarithmically equal spacing from $r = r_{\mathrm{in}}$ to $r =$100AU at $t = 0$.

The particle growth rate, $\dot{m}_{\mathrm{p},k}$, the radial drift velocity, $v_{r\mathrm{p}.k}$, the vertical settling velocity $v_{z\mathrm{p},k}$, and the escape velocity from the particle, $v_{\mathrm{esc},k}$, are given by:

$$\dot{m}_{\mathrm{p},k} = \pi a_{\mathrm{p},k}^2 \rho_{\mathrm{p}}(r_{\mathrm{p},k}) v_{\mathrm{rel,pp},k} \left(1 + \frac{v_{\mathrm{esc},k}^2}{v_{\mathrm{rel,pp},k}^2}\right) \quad (46)$$

$$v_{r\mathrm{p},k} = \frac{2\Omega_k t_{\mathrm{s},k}}{1+(\Omega_k t_{\mathrm{s},k})^2}\eta_k r_{\mathrm{p},k}\Omega_k + v_{r\mathrm{KH},k}, \quad (47)$$

$$v_{z\mathrm{p},k} = \frac{\Omega_k t_{\mathrm{s},k}}{1+\Omega_k t_{\mathrm{s},k}} z_{\mathrm{p},k}\Omega_k, \quad (48)$$

$$v_{\mathrm{esc},k} = \sqrt{\frac{2Gm_{\mathrm{p},k}}{a_{\mathrm{p},k}}}, \quad (49)$$

where $\eta_k = -\frac{1}{2}\left[\frac{c_\mathrm{s}^2}{r^2\Omega^2}\frac{\partial\log(\rho_\mathrm{m}T_\mathrm{m})}{\partial\log r}\right]_k$, and the particle radius $a_{\mathrm{p},k}$ is calculated by the particle internal density $\rho_{\mathrm{i},k}$ for a given mass $m_{\mathrm{p},k}$ as:

$$a_{\mathrm{p},k} = \left(\frac{3m_{\mathrm{p},k}}{4\pi\rho_{\mathrm{i},k}}\right)^{1/3}. \quad (50)$$

The particle stopping time $t_{\mathrm{s},k}$ (equation 64), and the relative particle-particle relative velocity $v_{\mathrm{rel,pp},k}$ (equation 66) are given, later. The second term of equation (47) represents the radial drift velocity due to the turbulence by Kelvin-Helmholtz instability in the subdisk. It is given by:

$$v_{r\mathrm{KH},k} = \begin{cases} \dfrac{2\alpha_{\mathrm{D},k}\rho_\mathrm{m}(r_{\mathrm{p},k})c_\mathrm{s}(r_{\mathrm{p},k})^2}{\Sigma_{\mathrm{p},k}\Omega_k\mathrm{Re}^*} & \text{for} \quad \rho_{\mathrm{pm},k} > \rho_{\mathrm{m},k}, \\ 0 & \text{for} \quad \rho_{\mathrm{pm},k} < \rho_{\mathrm{m},k}, \end{cases} \quad (51)$$

where $\mathrm{Re}^* = 55$ (Cuzzi et al., 1993) and the vertical diffusion parameter $\alpha_{\mathrm{D},k}$ is given in equation (73). Finally, the particle drift timescale and the particle growth timescale can be defined as:

$$\tau_{\mathrm{drift},k} = \frac{r_{\mathrm{p},k}}{|v_{r\mathrm{p},k}|}, \quad (52)$$

$$\tau_{\mathrm{grow},k} = \frac{m_{\mathrm{p},k}}{\dot{m}_{\mathrm{p},k}}. \quad (53)$$

When two superparticles $k$ and $k'$ become closer each other than $\max(H_k/2, H_{k'}/2)$, the superparticle $k'$ is merged into the superparticle $k$, where $m_{\mathrm{p},k} > m_{\mathrm{p},k'}$. The new superparticle $k$ is created at the center of gravity between $k$ and $k'$, in other words:

$$m_{\mathrm{p},k}^{\mathrm{new}} = \max(m_{\mathrm{p},k}, m_{\mathrm{p},k'}), \quad (54)$$

$$M_{\mathrm{p},k}^{\mathrm{new}} = M_{\mathrm{p},k} + M_{\mathrm{p},k'}, \quad (55)$$

$$r_{\mathrm{p},k}^{\mathrm{new}} = (M_{\mathrm{p},k}r_{\mathrm{p},k} + M_{\mathrm{p},k'}r_{\mathrm{p},k'})/M_{\mathrm{p},k}^{\mathrm{new}}, \quad (56)$$

$$k_+^{\mathrm{new}} = \max(k, k') + 1, \quad (57)$$

$$k_-^{\mathrm{new}} = \min(k, k') - 1. \quad (58)$$

Here, it is worth noting that the artificial shift of the boundaries between merging particles and inner and outer nearest particles by the equation 39 is not significant in the particular cases discussed in the present paper, since their merging always takes place when a superparticle drifts inward for a large distance to merge the inner superparticle, leaving the outer superparticle, in which solid particles did not yet grow enough to drift down.

*4.1.2. Internal Density and Stopping Time*

In the rocky region, where $T_{\rm m} > 180$ K, we assume: $\rho_{{\rm i},k} = \rho_{\rm rock} = 2\,{\rm g\,cm^{-3}}$. On the other hand, in the icy region, $T_{\rm m} < 150$ K, $\rho_{{\rm i},k}$ is given by:

$$\rho_{{\rm i},k} = \rho_{\rm ice} = \min[\,\max(\,\rho_{{\rm BCCA},k}, \rho_{{\rm coll},k}, \rho_{{\rm gas},k}, \rho_{{\rm G},k}\,),\ \rho_0]. \tag{59}$$

Here, $\rho_{{\rm BCCA},k}$, $\rho_{{\rm coll},k}$, $\rho_{{\rm gas},k}$, $\rho_{{\rm G},k}$, and $\rho_0$ are given below following Kataoka et al. (2013) and Okuzumi et al. (2012). In the intermediate temperature, where $150\,{\rm K} < T_{\rm m} < 180$ K, $\rho_{{\rm i},k}$ is linearly interpolated from $\rho_{\rm rock}$ and $\rho_{\rm ice}$.

First, when the impact energy of particle-particle collision, $E_{{\rm imp},k} = m_{{\rm p},k} v_{{\rm rel,pp},k}^2/4$, is less than the rolling energy of two contacting monomers, $E_{\rm roll} = 4.74 \times 10^{-9}$ erg, the particles grow by the hit-and-stick mode. This is called the BCCA particle and is consisted of many monomers with the radius $a_0 \sim 0.1\mu$m and the density $\rho_0 \sim 1\,{\rm g\,cm^{-3}}$. The internal density of the BCCA particle decreases with $m_{{\rm p},k}^{-1/2}$ as:

$$\rho_{{\rm BCCA},k} = \left(\frac{m_{{\rm p},k}}{m_0}\right)^{-\frac{1}{2}} \rho_0, \tag{60}$$

where $m_0 = 4\pi a_0^3 \rho_0/3$ is the monomer mass.

Second, when the impact energy exceeds the rolling energy, the particles are compressed by particle-particle collisions. The internal density in such collisional-compression phase

is expressed as (Suyama et al., 2008, 2012):

$$\rho_{{\rm coll},k} = \frac{1}{\sqrt{2}}\left(\frac{3}{5}\right)^{3/2}\left(\frac{m_0 v_{{\rm rel,pp},k}^2}{0.60 E_{\rm roll}}\right)^{3/10}\left(\frac{m_{{\rm p},k}}{m_0}\right)^{-1/5}\rho_0. \tag{61}$$

Third, when the gas pressure is dominant, the internal density is given by:

$$\rho_{{\rm gas},k} = \left(\frac{a_0^3 m_{{\rm p},k} v_{{\rm rel,pg},k}}{\pi a_{{\rm p},k}^2 t_{{\rm s},k} E_{{\rm roll},k}}\right)^{\frac{1}{3}} \rho_0, \tag{62}$$

where $v_{\mathrm{rel,pg},k}$ is the relative velocity between particle and gas, which is given later in equation (76).

Finally when the self gravitational force of the particle is dominant:

$$\rho_{\mathrm{G},k} = \left( \frac{G m_{\mathrm{p},k}^2 a_0^3}{\pi a_{\mathrm{p},k}^4 E_{\mathrm{roll}}} \right)^{\frac{1}{3}} \rho_0. \tag{63}$$

The particle stopping time is calculated as:

$$t_{\mathrm{s},k} = \frac{2 m_{\mathrm{p},k}}{\pi a_{\mathrm{p},k}^2 C_{\mathrm{D},k} \rho_k v_{\mathrm{rel,pg},k}}, \tag{64}$$

where

$$C_{\mathrm{D},k} = \begin{cases} \frac{8}{3} \frac{v_{\mathrm{th},k}}{v_{\mathrm{rel,pg},k}} & \text{for Epsetein Drag Regime,} \\ 6 \frac{v_{\mathrm{th},k}}{v_{\mathrm{rel,pg},k}} \frac{\lambda_{\mathrm{mfp},k}}{a_{\mathrm{p},k}} & \text{for Stokes Drag Regime,} \\ 0.44 & \text{for Newton Drag Regime .} \end{cases} \tag{65}$$

Here, $\rho_k$, $v_{\mathrm{th},k}$, and $\lambda_{\mathrm{mfp},k}$ are the gas density, the thermal velocity, and the mean free path of the gas molecule at $r = r_{\mathrm{p},k}, z = z_{\mathrm{p},k}$, respectively.

*4.1.3. Relative Velocity*

The particle-particle relative velocity, $v_{\mathrm{rel,pp},k}$, is given by the sum of Brownian motion $v_{\mathrm{B},k}$, radial drift difference $v_{r\mathrm{pp},k}$, azimuthal drift difference $v_{\phi\mathrm{pp},k}$, vertical settling difference $v_{z\mathrm{pp},k}$, turbulent velocity $v_{\mathrm{turb,pp},k}$, and viscous stirring velocity $v_{\mathrm{VS},k}$ as:

$$v_{\mathrm{rel,pp},k} = \sqrt{v_{\mathrm{B},k}^2 + v_{r\mathrm{pp},k}^2 + v_{\phi\mathrm{pp},k}^2 + v_{z\mathrm{pp},k}^2 + v_{\mathrm{turb,pp},k}^2 + v_{\mathrm{VS},k}^2}, \tag{66}$$

where:

$$v_{\mathrm{B},k} = \sqrt{\frac{16}{\pi} \frac{k_{\mathrm{B}} T_{\mathrm{m}}(r_{\mathrm{p},k})}{m_{\mathrm{p},k}}}, \tag{67}$$

$$v_{r\mathrm{pp},k} = \left( \frac{2\Omega_k t_{\mathrm{s},k}}{1 + (\Omega_k t_{\mathrm{s},k})^2} - \frac{2(0.5\Omega_k t_{\mathrm{s},k})}{1 + (0.5\Omega_k t_{\mathrm{s},k})^2} \right) \eta_k r_{\mathrm{p},k} \Omega_k, \tag{68}$$

$$v_{\phi\mathrm{pp},k} = -\left( \frac{(\Omega_k t_{\mathrm{s},k})^2}{1 + (\Omega_k t_{\mathrm{s},k})^2} - \frac{(0.5\Omega_k t_{\mathrm{s},k})^2}{1 + (0.5\Omega_k t_{\mathrm{s},k})^2} \right) \eta_k r_{\mathrm{p},k} \Omega_k, \tag{69}$$

and

$$v_{z\mathrm{pp},k} = \left( \frac{\Omega_k t_{\mathrm{s},k}}{1 + \Omega_k t_{\mathrm{s},k}} - \frac{0.5\Omega_k t_{\mathrm{s},k}}{1 + 0.5\Omega_k t_{\mathrm{s},k}} \right) z_{\mathrm{p},k} \Omega_k. \tag{70}$$

The turbulent velocity between two particles is (Ormel and Cuzzi, 2007):

$$v_{\mathrm{turb,pp},k} = \sqrt{\alpha_{\mathrm{D},k}} c_{\mathrm{s}}(r_{\mathrm{p},k}) \times \begin{cases} \mathrm{Re}_{\mathrm{t},k}^{\frac{1}{4}} \Omega_k |0.5 t_{\mathrm{s},k}|, & \text{for} \quad \Omega_k t_{\mathrm{s},k} < \Omega_k t_{\eta,k}, \\ \sqrt{2\Omega_k t_{\mathrm{s},k}} & \text{for} \quad \Omega_k t_{\eta,k} < \Omega_k t_{\mathrm{s},k} < 1, \\ \sqrt{\frac{1}{1+\Omega_k t_{\mathrm{s},k}} + \frac{1}{1+0.5\Omega_k t_{\mathrm{s},k}}} & \text{for} \quad 1 < \Omega_k t_{\mathrm{s},k}, \end{cases} \tag{71}$$

where non-dimensional turnover time of the smallest eddy is expressed as $\Omega_k t_{\eta,k} = \mathrm{Re}_{\mathrm{t},k}^{-1/2}$. This estimate indicates that the representative particle-particle velocity is comparable to the relative velocity between two different particles with $t_{\mathrm{s}}$ and $0.5t_{\mathrm{s}}$. This choice of factor 0.5 is consistent with the recent work by Sato et al. (2015).

The turbulent Reynolds number is given by:

$$\mathrm{Re}_{\mathrm{t},k} = \frac{2\alpha_{\mathrm{D},k} c_{\mathrm{s}}(r_{\mathrm{p},k})^2}{\Omega_k \lambda_{\mathrm{mfp},k} v_{\mathrm{th},k}}, \tag{72}$$

where the viscous parameter $\alpha_{\mathrm{D},k}$ is given by:

$$\alpha_{\mathrm{D},k} = \begin{cases} \bar{\alpha} & \text{for turbulent region,} \\ 0 & \text{for } \rho_{\mathrm{pm},k} < \rho_{\mathrm{m},k} \text{ in quiet area,} \\ 0.19 \left( \dfrac{\eta_k r_{\mathrm{p},k}}{H(r_{\mathrm{p},k})} \right)^2 \min(\Omega_k t_{\mathrm{s},k}, 1) & \text{for } \rho_{\mathrm{pm},k} > \rho_{\mathrm{m},k} \text{ in quiet area,} \end{cases} \tag{73}$$

depending on the turbulent state. Here, we adopt the criterion of $\rho_{\mathrm{pm},k} > \rho_{\mathrm{m},k}$ for the activation of Kelvin–Helmholtz instability in the quiet area. We use the formula of viscous parameter by Takeuchi et al. (2012) slightly modified as scaled as $\min(\Omega_k t_{\mathrm{s},k}, 1)$ instead of $\Omega_k t_{\mathrm{s},k}$. Here, we take into account the mechanical decoupling between gas and particles.

After the gravitational instability takes place as described in §4.1.5, the particles in the subdisk are gravitationally perturbed by massive bodies (planetesimals). The equilibrium eccentricity is determined by the competition of the stirring from the massive bodies and the damping from gas drag. According to Kobayashi et al. (2010), the equilibrium eccentricity $e_{\mathrm{VS},k}$ is given by:

$$e_{\mathrm{VS},k} = \left( \frac{6 m_{\mathrm{p},k}}{2^{\frac{1}{3}} \eta_k \pi^2 a_{\mathrm{p},k}^2 C_{\mathrm{D},k} \rho_{\mathrm{m}}(r_{\mathrm{p},k}) r_{\mathrm{p},k}} \right)^{\frac{1}{4}} \left( \frac{m_{\mathrm{p},k}}{3 M_\odot} \right)^{\frac{5}{12}}. \tag{74}$$

Here, we assume $e_{\rm VS} \simeq 2i_{\rm VS}$. The viscous stirring velocity is given by:

$$v_{{\rm VS},k} = \left(e_{{\rm VS},k}^2 + i_{{\rm VS},k}^2\right)^{\frac{1}{2}} r_{{\rm p},k}\Omega_k = \sqrt{\frac{5}{4}} e_{{\rm VS},k} r_{{\rm p},k}\Omega_k. \quad (75)$$

On the other hand, the particle-gas relative velocity, $v_{{\rm rel,pg},k}$, is calculated as follows:

$$v_{{\rm rel,pg},k} = \sqrt{v_{{\rm B},k}^2 + v_{r{\rm pg},k}^2 + v_{\phi{\rm pg},k}^2 + v_{z{\rm pg},k}^2 + v_{{\rm turb,pg},k}^2 + v_{{\rm VS},k}^2}, \quad (76)$$

where:

$$v_{r{\rm pg},k} = \frac{2\Omega_k t_{{\rm s},k}}{1+(\Omega_k t_{{\rm s},k})^2}\eta_k r_{{\rm p},k}\Omega_k \quad (77)$$

$$v_{\phi{\rm pg},k} = -\frac{(\Omega_k t_{{\rm s},k})^2}{1+(\Omega_k t_{{\rm s},k})^2}\eta_k r_{{\rm p},k}\Omega_k \quad (78)$$

and

$$v_{z{\rm pg},k} = \frac{\Omega_k t_{{\rm s},k}}{1+\Omega_k t_{{\rm s},k}} z_{{\rm p},k}\Omega_k. \quad (79)$$

The turbulent velocity between particle and gas is (Ormel and Cuzzi, 2007):

$$v_{{\rm turb,pg},k} = \sqrt{\alpha_{{\rm D},k}} c_{\rm s}(r_{{\rm p},k}) \times \begin{cases} {\rm Re}_{{\rm t},k}^{\frac{1}{4}}\Omega_k |t_{{\rm s},k}|, & \text{for} \quad \Omega_k t_{{\rm s},k} < \Omega_k t_{\eta,k}, \\ \sqrt{3\Omega_k t_{{\rm s},k}} & \text{for} \quad \Omega_k t_{\eta,k} < \Omega_k t_{{\rm s},k} < 1, \\ \sqrt{\frac{1}{1+\Omega_k t_{{\rm s},k}} + 1} & \text{for} \quad 1 < \Omega_k t_{{\rm s},k}. \end{cases} \quad (80)$$

*4.1.4. Particle Fragmentation*

When the high velocity collision occurs, the particle is fragmented rather than coagulate. The fragmentation velocity, $v_{{\rm frag},k}$, is set $60\,{\rm m\,s^{-1}}$ in the icy region (Wada et al., 2009). In the rocky region, it becomes the size dependent value as (Ormel and Okuzumi, 2013):

$$v_{{\rm frag},k} = \left[8\left(q_{\rm s} R_{{\rm C1},k}^{\frac{9\mu_{\rm p}}{3-2\phi_{\rm p}}} + q_{\rm g} R_{{\rm C1},k}^{3\mu_{\rm p}}\right)\right]^{\frac{1}{3\mu_{\rm p}}}, \quad (81)$$

where

$$R_{{\rm C1},k} = \left(\frac{3}{2\pi}\frac{m_{{\rm p},k}}{1[{\rm g\ cm^{-3}}]}\right)^{\frac{1}{3}}, \quad (82)$$

and $q_s = 500$, $q_g = 10^{-4}$, $\mu_p = 0.4$, and $\phi_p = 7$ are the fitting parameter in cgs unit, which is given by Stewart and Leinhardt (2009) as a regime of weak rocks. When the particle fragmentation occurs, we simply reduce the particle mass so that the $v_{\mathrm{rel,pp},k}$ becomes $v_{\mathrm{frag},k}$.

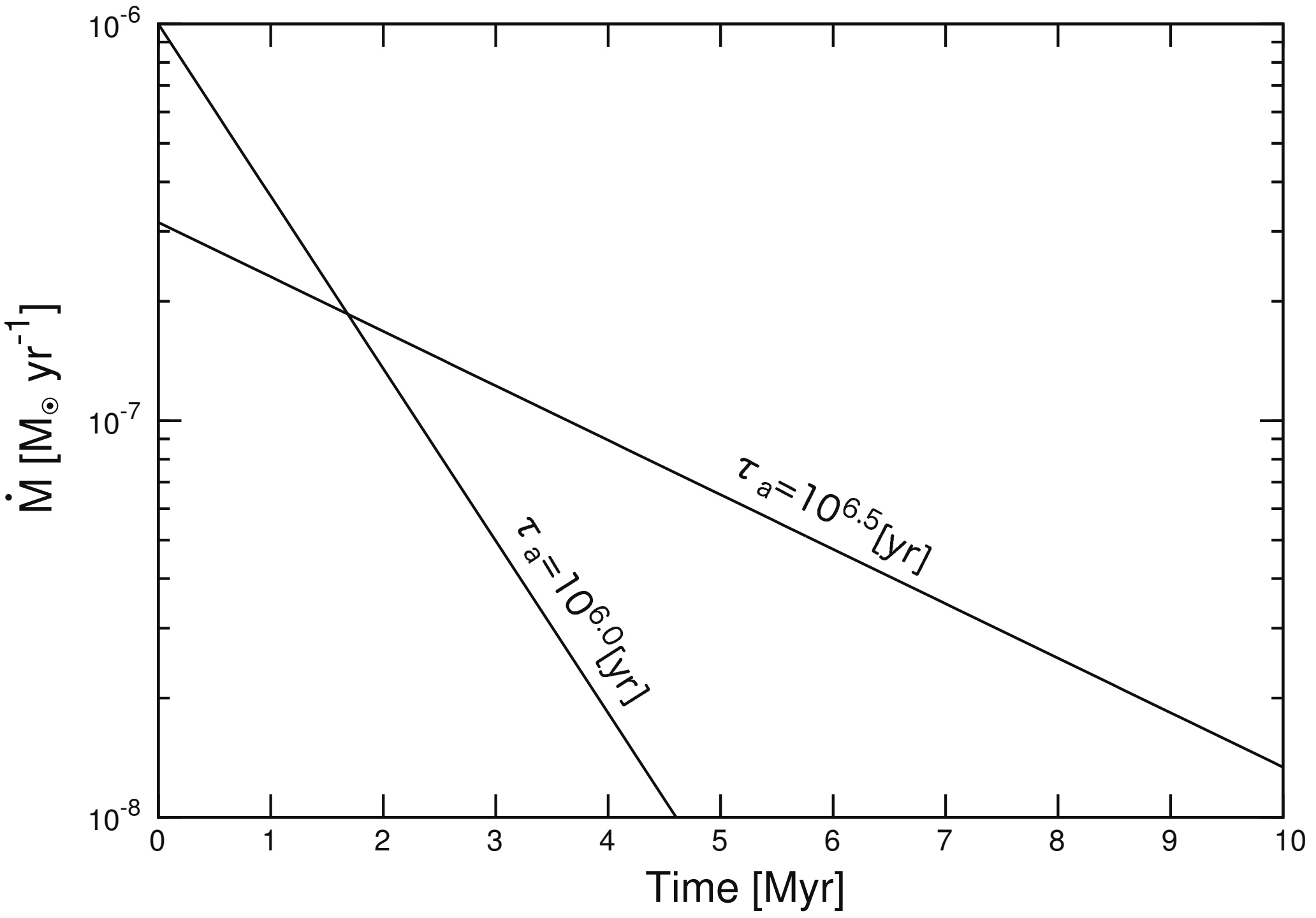

Figure 7: Probable time variation of the accretion rate, $\dot{M}$. Here, we assume the mass of the central star to be $1\mathrm{M}_\odot$ and an exponential decrease with decay times: $\tau_a = 10^{6.0}$ yr and $10^{6.5}$ yr.

*4.1.5. Formation (Gravitational Instability) and Growth of Planetesimals*

The onset condition for gravitational instability is given by Yamoto and Sekiya (2004):

$$\rho_{\mathrm{GI},k} = 0.78 \frac{M_*}{r_{\mathrm{p},k}^3}, \tag{83}$$

and the initial planetesimal mass, $m_{\mathrm{plt},k}$ is estimated using the most unstable wavelength as:

$$m_{\mathrm{plt},k} = \lambda_{\mathrm{GI},k}^2 \rho_{\mathrm{pm},k} z_{\mathrm{p},k}, \tag{84}$$

where

$$\lambda_{\mathrm{GI},k} = \frac{2\pi z_{\mathrm{p},k}}{0.26}. \tag{85}$$

When the particle density at the midplane, $\rho_{\mathrm{pm},k}$, becomes larger than the above critical density, $\rho_{\mathrm{pm},k} > \rho_{\mathrm{GI},k}$, the planetesimal with mass of $m_{\mathrm{plt},k}$ is created by gravitational instability. We freeze the evolution of pebble after that. Although the evolution of the planetesimals after the gravitational instability is almost beyond the scope of the present paper, we further study in order to know whether or not the planetesimals could grow to the Earth's mass ($m_\oplus$) in the environment, as follows. The growth rate of planetesimals after the gravitational instability is calculated as

$$\begin{aligned} \frac{dm_{\mathrm{plt},k}}{dt} &= \frac{1}{1.3\times 10^5}\left(\frac{\Sigma(r_{\mathrm{p},k})}{2400\,\mathrm{g\,cm^{-2}}}\right)^{2/5}\left(\frac{m_{\mathrm{plt},k}}{M_\oplus}\right)^{2/3}\left(\frac{\rho_{\mathrm{i},k}}{2\,\mathrm{g\,cm^{-3}}}\right)^{-3/5} \\ &\quad\times\left(\frac{r_{\mathrm{p},k}}{1\,\mathrm{AU}}\right)^{-3/5}\left(\frac{\Sigma_{\mathrm{p},k}}{10\,\mathrm{g\,cm^{-2}}}\right)\left(\frac{m_{\mathrm{p},k}}{10^{18}\,\mathrm{g}}\right)^{-2/15}\left(\frac{M_\oplus}{1\,\mathrm{yr}}\right), \end{aligned} \tag{86}$$

(Kokubo and Ida, 2012). It is consistent with the equation 16 of Kobayashi et al. (2010) within factor of 2 when we use the equilibrium eccentricity given by equation 74. The migration of planetesimals is generally negligible for the case of $m_{\mathrm{plt}} < M_\oplus$ in the tandem disk.

### *4.1.6. Initial Condition and Iteration*

A new star is formed from the gravitational collapse of a dense molecular cloud. Since its free fall time scale is $\sim 10^6$ yrs, it is likely to take 1-2 Myr to establish a steady-state accretion disk around the newly born star. As can be seen in figure 7, mass accretion rate decrease down to $\sim 10^{-7} M_\odot\,\mathrm{yr}^{-1}$, for the time scale of $\tau_a = 10^{6.0}\,\mathrm{yr}$. Therefore, we use the disk structure with $\dot{M} = 10^{-7} M_\odot\,\mathrm{yr}^{-1}$ in the present section.

In the first 1-2 Myr of the star formation, the gas inflow is expected to be fully turbulent and the direct accretion from the z-direction is significant compared with the accretion flux from the midplane, inside of the disk. In such a violently turbulent disk, particles are likely to be well mixed with gas and not to grow significantly. Therefore, we started the particle growth calculation from various radius $r_{\mathrm{s}}$, assuming that the initial distribution of particle is the same as that of gas, in other words, $H_{\mathrm{p},k} = H(r_{\mathrm{s},k})$ and that $m_{\mathrm{p}} = m_0$.

Since the equations from (47) through (82) are related each other, they are solved iteratively to determine the various values: $v_{r\mathrm{p},k}$, $v_{z\mathrm{p},k}$, $t_{\mathrm{s},k}$, $H_{\mathrm{pe},k}$, $a_{\mathrm{p},k}$, $v_{\mathrm{rel,pp},k}$, $v_{r\mathrm{pp},k}$, $v_{\phi\mathrm{pp},k}$, $v_{z\mathrm{pp},k}$, $v_{\mathrm{turb,pp},k}$, $v_{\mathrm{rel,pg},k}$, $v_{r\mathrm{pg},k}$, $v_{\phi\mathrm{pg},k}$, $v_{z\mathrm{pg},k}$, $v_{\mathrm{turb,pg},k}$, $v_{\mathrm{VS},k}$, $v_{\mathrm{frag},k}$, $v_{\mathrm{esc},k}$, $\rho_{\mathrm{i},k}$, $\rho_{\mathrm{coll},k}$, $\rho_{\mathrm{gas},k}$, $\rho_{\mathrm{G},k}$, $E_{\mathrm{imp},k}$, $\mathrm{Re}_{\mathrm{t},k}$, $\alpha_{\mathrm{D},k}$, and $C_{\mathrm{D},k}$. After obtaining these values, the evolution equations from (40) through (43) are calculated to obtain new coordinates.

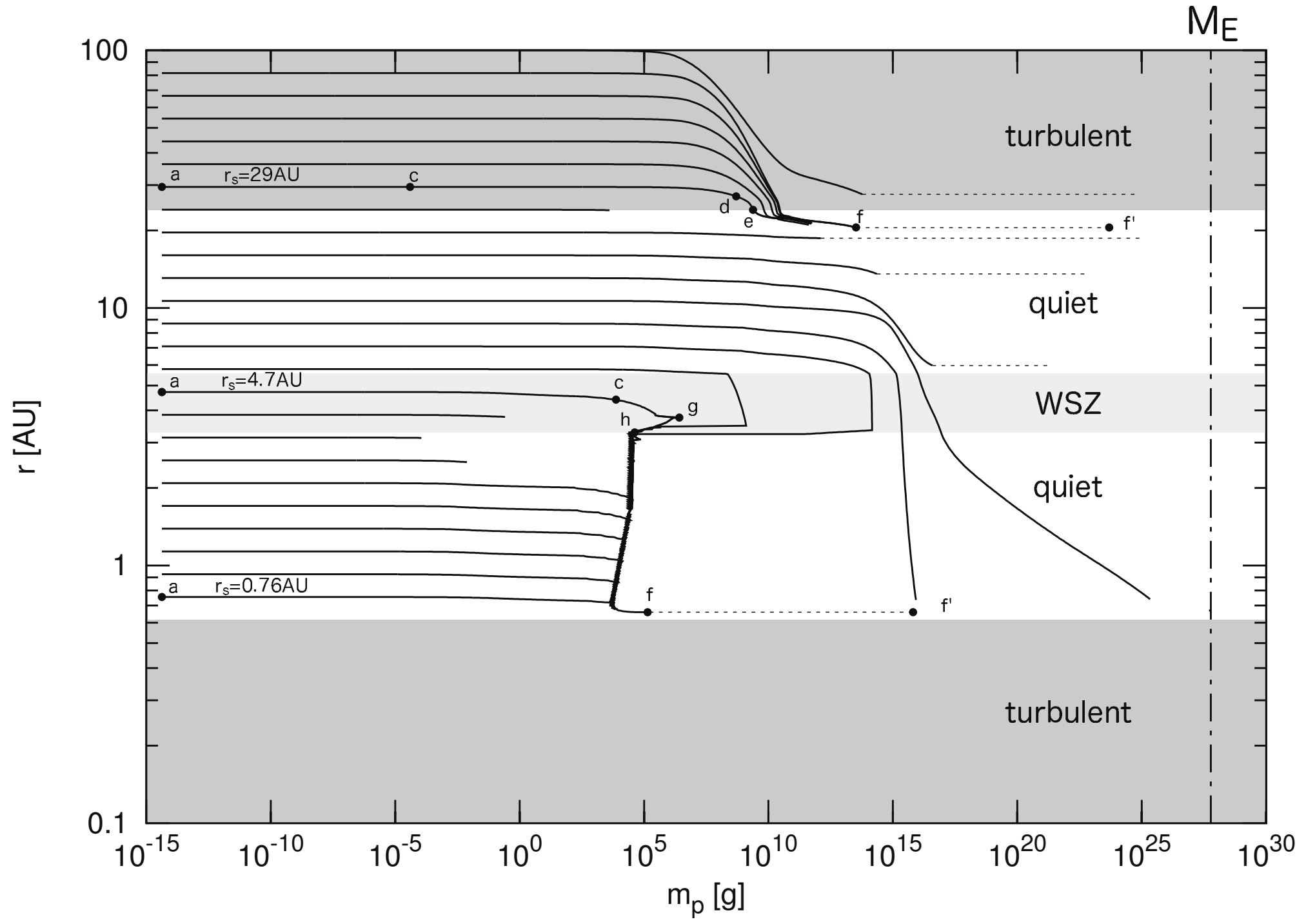

Figure 8: The growth of particles in a mass-radius diagram for the disk with $\dot{M} = 10^{-7.0}\,\mathrm{M}_\odot\,\mathrm{yr}^{-1}$. Icy planetesimals undergo runaway growth to $m_\mathrm{p} \sim 10^{7-8}$ g in OTR and drift inward to the outer MRI front ($r = r_\mathrm{out}$). Solid particles rapidly grow in the quiet area with higher particle density in the turbulent-free environment leading to gravitational instability to form icy planetesimals at the outer MRI front ($r = r_\mathrm{out}$). Rocky planetesimals are also formed at the inner MRI front ($r = r_\mathrm{in}$) of the quiet area, where they are accumulated around the pressure maximum. The letters denotes the epoch of the growth, see text in detail. Dashed lines represent growth through gravitational instability. We stop the calculation at $t = 10^6$ yrs or the onset of gravitational instability.

## *4.2. Planet Formation at the Outer MRI Front ($r = r_\mathrm{out}$)*

Icy particles starting from outside of the outer MRI front (i.e., $r_\mathrm{s} > r_\mathrm{out}$), grow through mutual collisions with low relative velocity less than 1cm/s. The resultant aggregations become porous with low density down to $10^{-5}\,\mathrm{g\,cm^{-3}}$(Okuzumi et al., 2012). Figures 8 and 9 show the evolution of a porous aggregation in mass-density diagram and time-mass diagram, starting from the point of $r = 29$ AU and $z = H(r_\mathrm{s} = 29\,\mathrm{AU})$ in the steady-state disk with $\dot{M} = 10^{-7.0}\,\mathrm{M}_\odot\,\mathrm{yr}^{-1}$. First, the internal density ($\rho_\mathrm{i}$) of the particle decreases since an aggregation grows ($a$ to $c$) as a ballistic cluster-cluster aggregate (BCCA). However, after the compression energy due to ram pressure dominates the rolling energy of two contacting monomers, it gradually increases ($c$ to $d$).

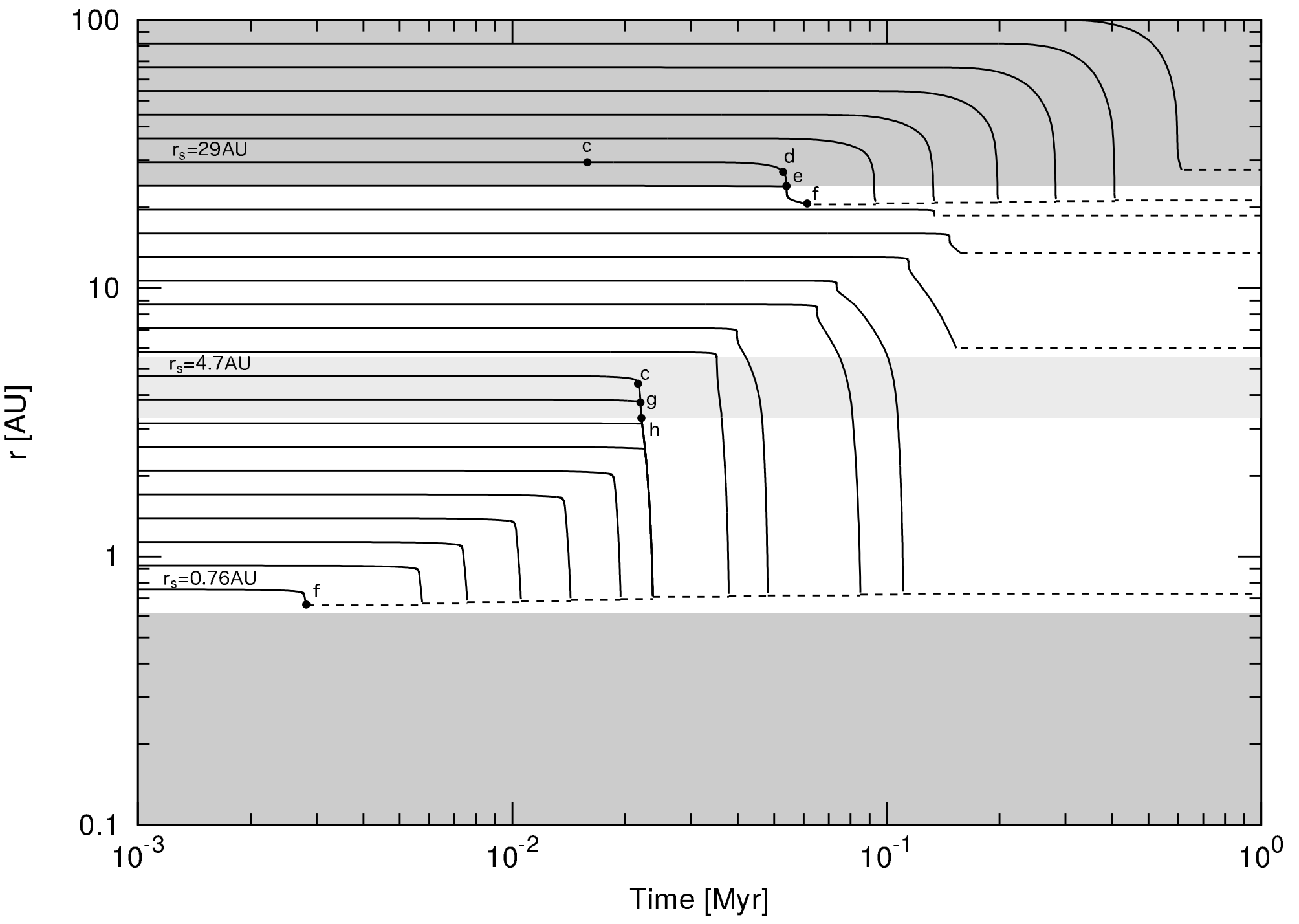

Figure 9: The same diagram but in a time-radius plane for the disk with $\dot{M} = 10^{-7.0}\,\mathrm{M}_\odot\,\mathrm{yr}^{-1}$. The superpartciles starts drifting inward one by one to merge with the inner superpaticles at the inner and outer MRI fronts, when they reach the drift barrier (figure 10). The letter $m$ represents the merging events among superparticles.

As the aggregate mass increases, the radial drift timescale becomes shorter than 30 times of growth timescale. This happens at point $d$, when the aggregate mass reaches $m_\mathrm{p} \sim 10^8$ g in the tandem disk (figures 8 and 9). Then, the aggregates drift ($d$ to $e$) into the quiet area (QA: $r = r_\mathrm{out}$), and sink down towards the disk midplane to form a sub-disk of pebbles with the thickness of $H_\mathrm{p}$ because of the lack of turbulence, as shown in figures 1 and 11 (Nakagawa et al., 1986; Cuzzi et al., 1993; Schrapler and Henning, 2004). Their sinking motion towards the midplane terminates when the density of the solid particles dominates over that of gas, i.e., $\rho_\mathrm{pm} > \rho_\mathrm{m}$, because of the shear tur-bulence driven by Kelvin-Helmholtz instability; The particle tends to rotate in Kepler velocity, $v_K$, while gas velocity is lower than $v_K$ by $\eta v_K$ ($\eta \sim 10^{-3}$) due to the pressure gradient in the disk (Youdin and Shu, 2002; Schrapler and Henning, 2004).

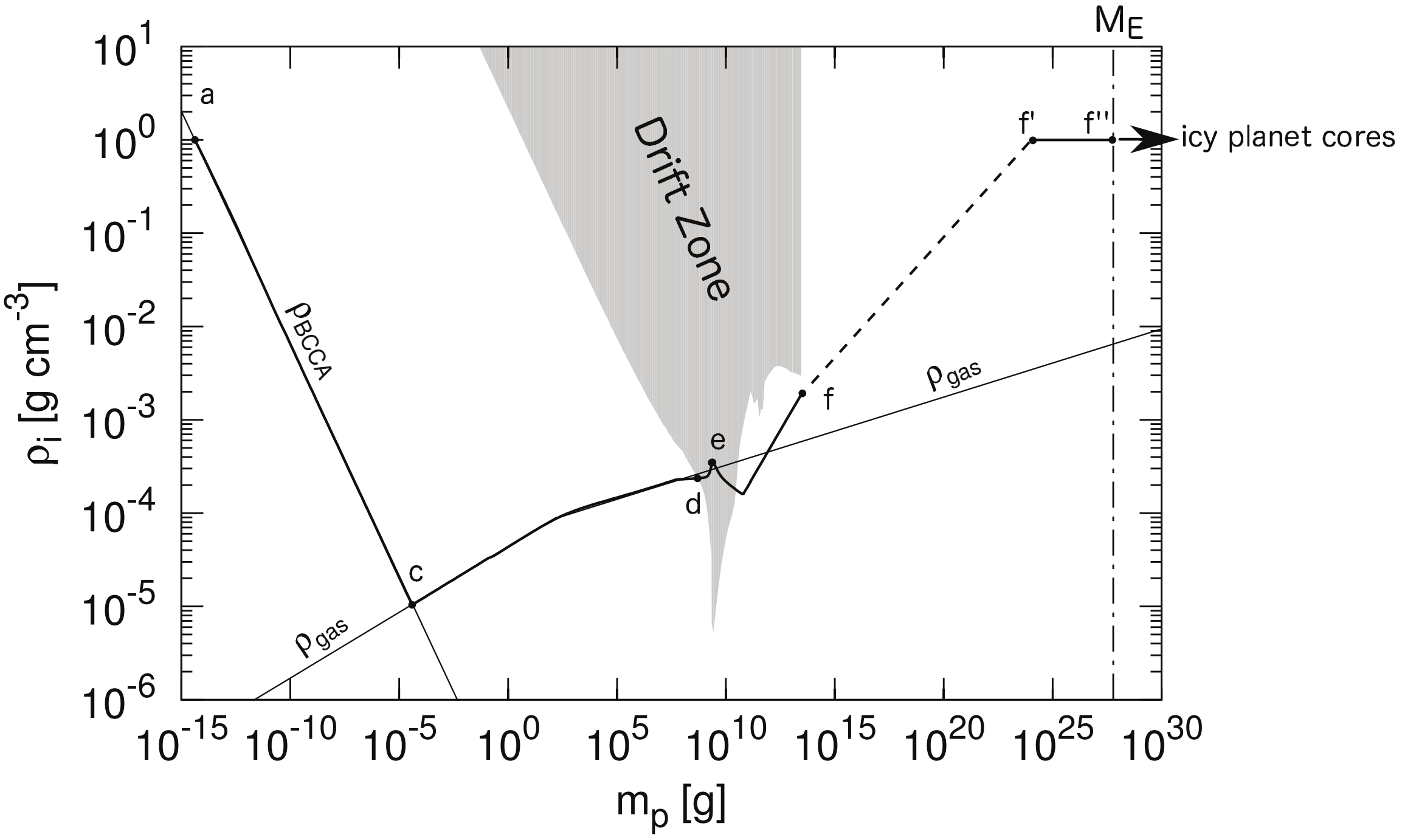

Figure 10: The growth history of the particle starting from $r_s = 29$AU shown in $m_p$-$\rho_i$ diagram. Dashed line represents the growth by gravitational instability. The hatched area represents the "drift zone", where $30\,\tau_{grow} > \tau_{drift}$. The letters attached in the figures denotes the epoch of the growth, see text in detail.

At this stage, the particles are decoupled with gas, in other words, $t_s\Omega > 1$. Cuzzi et al. (1993) studied the z-distribution of particles in such a decoupled case both analytically and numerically, and found that the scale height of particles of pebbles layer (subdisk) due to Kelvin-Helmholtz instability is much shorter than that of turbulent layer, in other words, Richardson number can be arbitrary small, depending on the situation, $J = \frac{g}{\rho}\frac{d\rho/dz}{(dU/dz)^2} \ll 1/4$. In fact, Chandrasekhar (1961) specially mentioned that $J \geq 1/4$ is a "necessary condition for stability." The subdisk, composed of particles decoupled with gas, can become very thin, as the particles grow and become more de-coupled. They eventually can undergo gravitational instability in spite of the turbulence around the midplane (figures 8 and 9). In fact, such a solution was obtained by Yamoto and Sekiya (2004), though they used it in a different context. In other words, the subdisk of pebbles can undergo gravitational instability without considerable particle enhancement, when the pebbles inside it are large enough to be decoupled with gas, even if the Kelvin-Helmholtz instability takes place around midplane. The previous authors (Sekiya, 1998; Youdin and Shu, 2002; Schrapler and Henning, 2004) unfortunately neglected this possibility, since they consider only the case that pebbles are well coupled with gas.

Figures 8 and 9 show the results of the particle growth calculation in a disk with $\dot{M} = 10^{-7.0}\,\mathrm{M}_\odot\,\mathrm{yr}^{-1}$. The planetesimals further grow through the accretion of pebbles to eventually reach an Earth mass $M_\oplus$ less than 1 Myr (Ormel and Klahr, 2010; Lambrechts and Johansen, 2012). For example, the planetesimals formed in the superparticles with $r_\mathrm{s} = 29$ AU reaches to $M_\oplus$ at $9 \times 10^5$ yr, as can be seen in figure 13. The pebbles with mass of $\sim 10^{10}$ g are continuously supplied from the outer turbulent region, as can be seen in figure 12.

### *4.3. Planet Formation around the Inner MRI Front ($r = r_\mathrm{in}$)*

The particles starting from the region of the water sublimation zone $r \sim r_\mathrm{WSZ}$ grow through mutual collisions. Entrapped water molecules begin to sublimate at the surfaces of the aggregates. The inclusions, such as silicate and organic matter, are likely packed together by surface tension of water; a liquid water phase likely takes place in the huge icy aggregations ($g \to h$). In the WSZ, alteration of nano-sized silica particles to clay minerals occurs upon reaction with residual water molecules, resulting in a net increase of the density ($\sim 2$ g cm$^{-3}$) of the newly transformed aggregates of clay particles (pebbles).

The aggregates continue to drift inward in the sub-disk of pebbles along the midplane, fragmented further by the weak turbulence due to the Kelvin–Helmholtz instability in the sub-disk of pebbles (figures 8 and 9). The inward drift terminates at the inner MRI front ($r = r_\mathrm{in}$). Since the radial pressure gradient turns to positive at the inner MRI front, a significant accumulation of particles takes place at the front (Kato et al., 2010). Figure 12 shows the increase in the solid-mass component mass $M_\mathrm{p}$ around $r = r_\mathrm{in}$. As the particles accumulate and continue to grow through the accretion of pebbles (Ormel and Klahr, 2010; Lambrechts and Johansen, 2012), they eventually undergo gravitational instability to become planetesimals. A planetesimals reached the Earth mass $M_\oplus$ at $6 \times 10^3$ yrs. The pebbles with the mass of $\sim 10^4$ g are continuously supplied from the outer areas through particle drift (figure 12). Since the midplane temperature of the inner MRI front reaches 1000 K, the rocky planetesimals loose almost all of their volatile components.

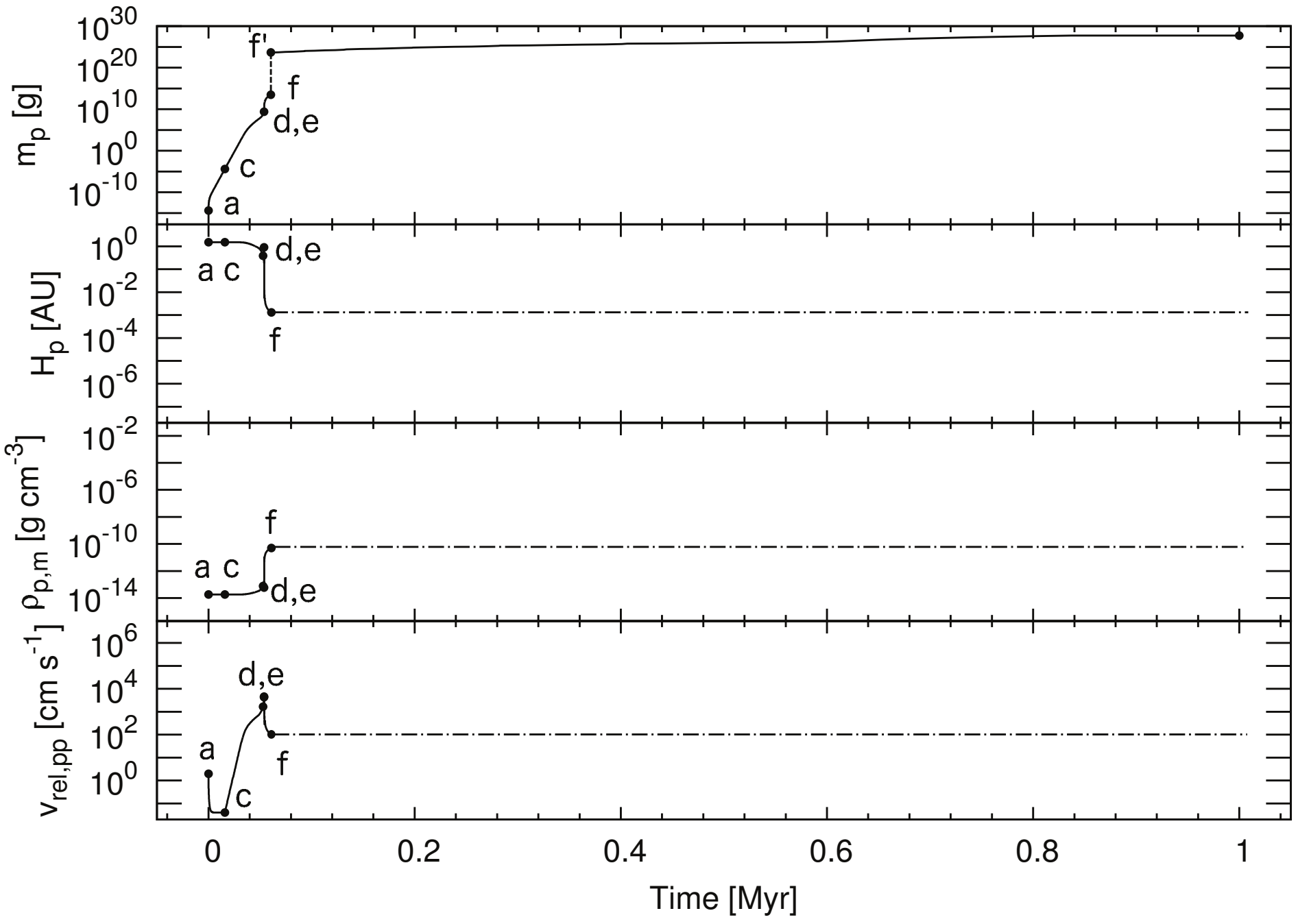

Figure 11: The time evolution of particle mass ($m_{\rm p}$), pebbles sub-disk thickness ($H_{\rm p}$), particle density $\rho_{\rm p,m}$ at the midplane, and relative velocity $v_{\rm pp,rel}$ for the particle starting from $r_{\rm s} = 29$ AU. The letters attached in the figures denotes the epoch of the growth, see text in detail.

## 5. Discussion and Summary

We have constructed a theoretical framework to discuss planet formation in the accretion disk around a young stellar object. We consider the on/off condition of Magneto-Rotational Instability through calculating the degree of ionization in the midplane of the gas disk, finding that the accretion disk has a quiet (i.e., non-turbulent) area in the MRI suppressed region (MSR). The quiet area is sandwiched between the outer and inner turbulent regions.

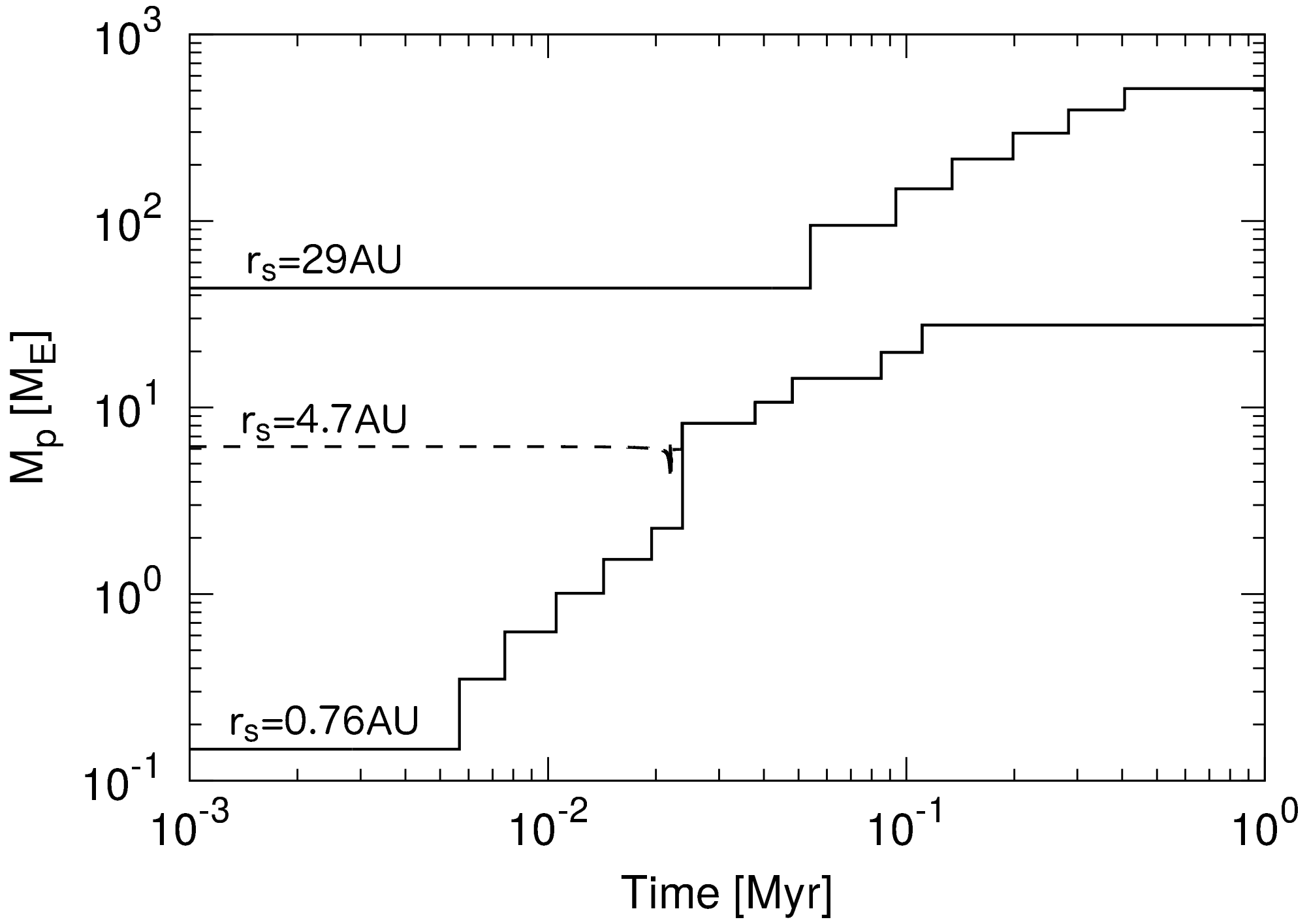

Figure 12: The total solid mass $M_{\rm p}$ included in the planetesimal formation (superparticles starting from $r_{\rm s} = 0.76,\ 4.7$, and 29 AU) takes place through the gravitational instability.

Solid particles in the gas disk drift inward and accumulate at the boundaries of the quiet area (the outer and inner MRI fronts). At the boundaries, the particles are accumulated both in the radial and the vertical directions and eventually undergo gravitational instability to form planetesimals. At least several planetesimals are likely to grow to $M_\oplus$ both the inner MRI fronts (for $6 \times 10^3$ yrs) and the outer MRI fronts (for $9 \times 10^5$ yrs). In other words, **Assumption A** of the standard model is therefore supported by our theoretical framework.

The solid particles, which accumulate at the two MRI fronts, settle down towards the midplane to form a sub-disk of pebbles. In the sub-disk, the density, $\rho_{\rm p}$, becomes so high at both the inner and outer MRI fronts that they undergo gravitational instability to form planetesimals. The radial concentration of solid particles is the key to overcoming the difficulties earlier discussed in the standard model, particularly in regards to the two following aspects. First, the radial drift of the pebbles assists the sub-disk to undergo gravitational instability (**Assumption B**), as discussed in subsections 4.2 and 4.3, even though Kelvin-Helmholtz instability takes place in the midplane (Youdin and Shu, 2002). Second, the growth of planetesimals are enhanced by a higher density of pebbles and by gas drag (so called "pebbles accretion":**Assumption D**; Ormel and Klahr, 2010; Lambrechts and Johansen, 2012). While **Assumption C** of the standard model is determined to be invalid here, our theoretical framework supports **Assumptions B** and **D**, as described above.

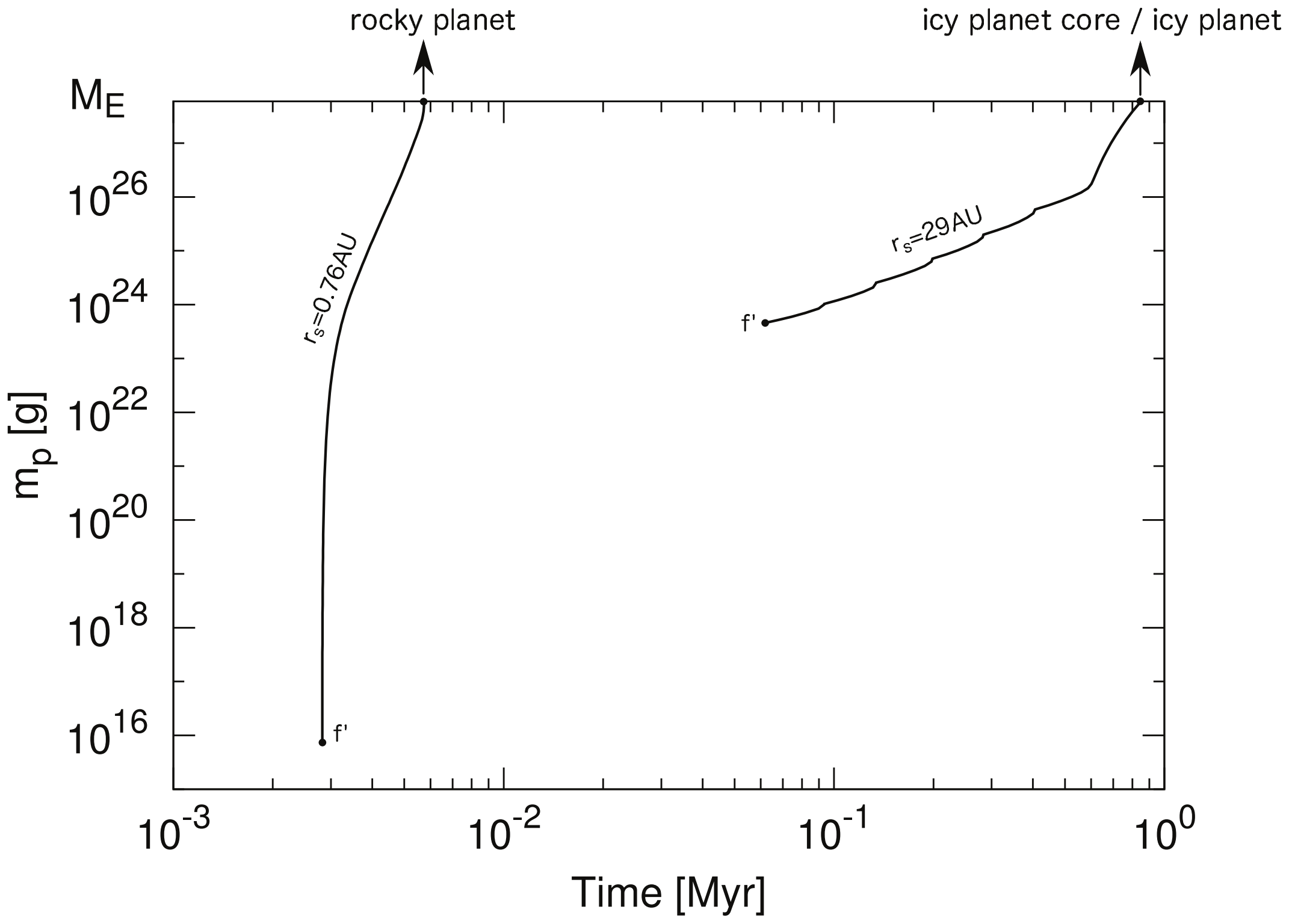

Figure 13: The growth of planetesimals in suparpartcles that started at $r_{\rm s}$ =0.76 and 29 AU.

The accretion rate, $\dot{M}$ onto a young stellar object is likely to decrease with time, as shown in figure 7, where we assume the mass of the central star to be $1\mathrm{M}_{\odot}$ and an exponential decrease with decay times: $\tau_{\rm a} = 10^6\,\mathrm{yr}$ and $3 \times 10^6\,\mathrm{yr}$. We find that planet cores are formed both in the outer and inner MRI fronts in the range of $\dot{M} = 10^{-8.0} - 10^{-6.5}\,\mathrm{M}_{\odot}\,\mathrm{yr}^{-1}$, in other words, the tandem planet formation robustly takes place in the accretion disk surrounding a young stellar object

The location of the MRI fronts depend on the accretion rate $\dot{M}$, as can be seen in figure 16. The radius $r_{out}$ of the outer MRI front locates at 60 AU for $\dot{M} = 10^{-6.5}M_\odot\,yr^{-1}$ and at 8 AU for $\dot{M} = 10^{-8.0}M_\odot\,yr^{-1}$, while that $r_{in}$ of inner MRI front locates at 1AU for $\dot{M} = 10^{-6.5}M_\odot\,yr^{-1}$ and at 0.2 AU for $\dot{M} = 10^{-8.0}M_\odot\,yr^{-1}$. This range of $r_{out}$ and $r_{in}$ suggests that there is no formation of planetismals in the range of 1-8 AU (figure 16). This lack of the planetesimal formation may correspond to the gap in the solid component distribution in the solar system around 2-4 AU, which reflects the fact that the mass of Mars is $0.1M_\oplus$ and the total mass in the asteroid belt is as low as $10^{-3}M_\oplus$ (Weidenschilling, 1977b). Such a gap in particle distribution is the key to reproducing a planetary system like our solar system. In fact, the major motivation of so called "Grand Tack" model (Walsh et al., 2011; Walsh and Morbidelli, 2011) was the formation of such a gap in the plenetesimal distribution by tacking of gas giants, Jupiter and Saturn. They assumed that Jupiter is first formed and migrates inward down to 1.5-2 AU. Then, Saturn grows and migrates as well to be locked in a 2:3 resonance with Jupiter.

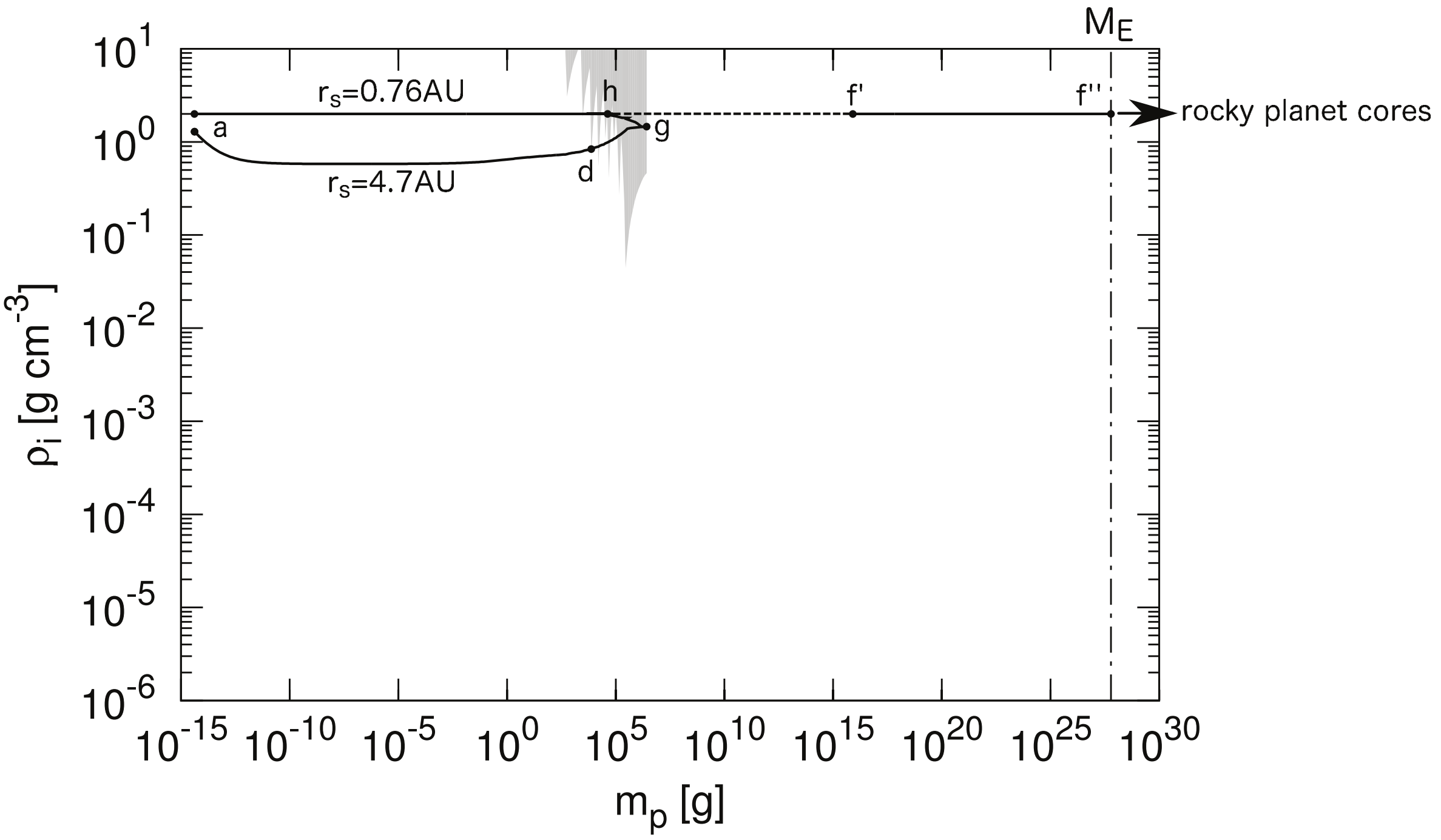

Figure 14: The growth history of particles starting at $r_s = 4.7$ AU, and $r_s = 0.76$ AU. The former is merged into latter at $t =$ and $m_p = 0.025$ Myrs.

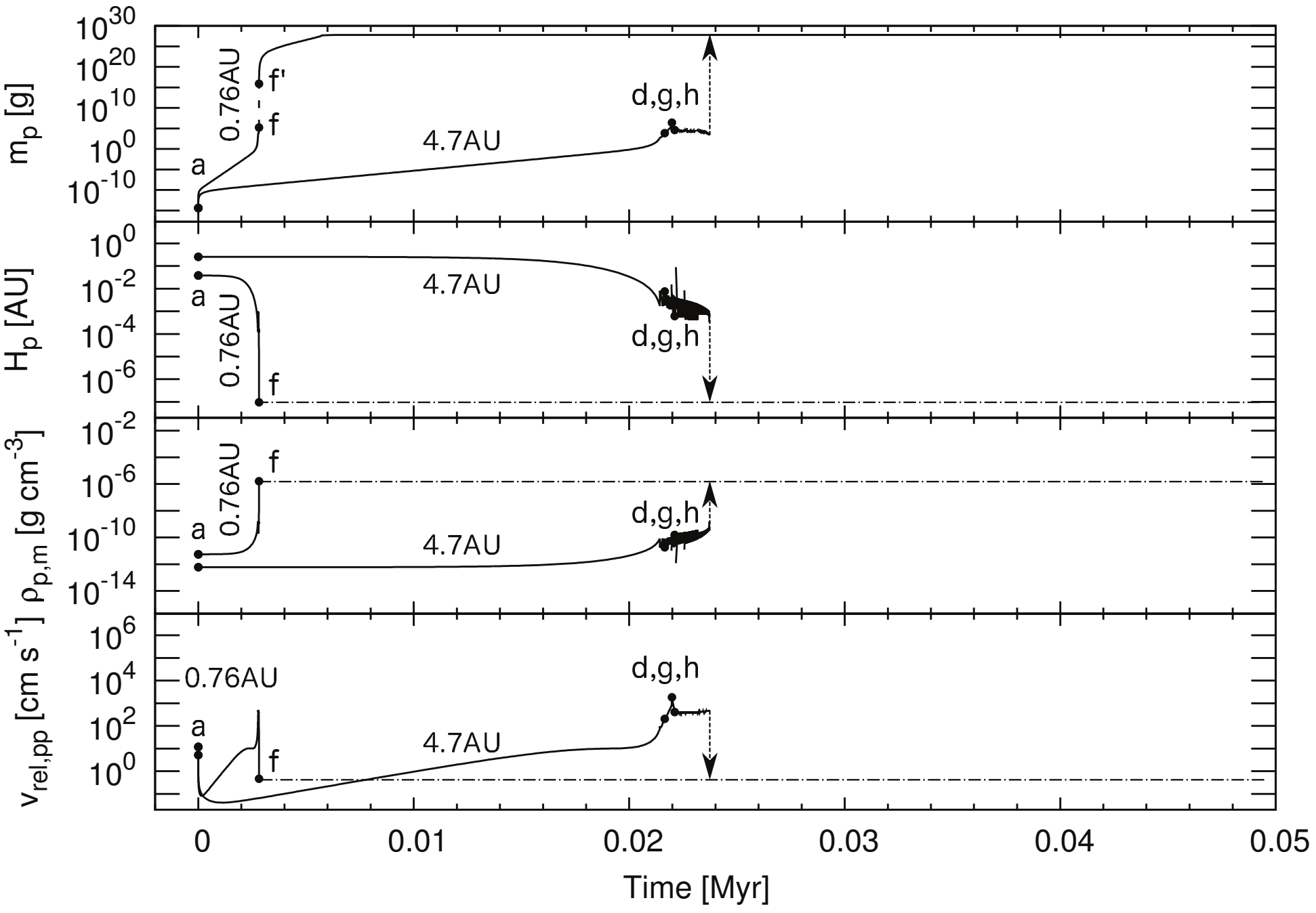

Figure 15: The time evolution of particle mass ($m_{\rm p}$), sub-disk thickness ($H_{\rm p}$), particle density $\rho_{\rm p,m}$ at the midplane, and relative velocity $v_{\rm rel,pp}$ for the super-particle stating from $r_{\rm s} = 4.7$ AU and $r_{\rm s} = 0.76$ AU.

The resonated pair of the planets migrates outward back to 5 and 7 AU. Gravitational scattering through Jupiter influences the inward migration of planetesimals and their eventual coalescence into a narrow torus around 1 AU.

On the other hand, the tandem planet formation naturally explains such a gap at 2-8 AU due to the tandem structure (i.e. two distinct sites) of the accretion disk during planetesimal formation (figure 16). It is worth noting that the model also explains the deficit of the solid component in the inner region than 0.7 AU, where "Grand Tack Model" has a difficulty to clean-up. In the tandem planet formation regime, on the other hand, the gas and therefore solid particles in the inner part ($r < 0.1$AU) of the disk with a low mass loss rate $\sim 10^{-8}\,{\rm M_\odot\,yr^{-1}}$ is likely lost, while that in the outer part ($r >$ several AU) still remains: The photoevaporation of gas from the disk can be as high as $\sim 10^{-8}\,{\rm M_\odot\,yr^{-1}}$ in particular if irradiated from extreme ultraviolet light or X-rays (in particular, Alexander et al., 2006; Owen et al., 2010), the disk with $\dot{M} \sim 10^{-8}\,\mathrm{M_\odot\,yr^{-1}}$ is considerably affected by the photoevapolation.

Furthermore, in tandem planet formation, pebbles are continuously supplied from the outer regions to the two formation sites (inner and outer MRI fronts; figures 8 and 9). This nature of the model helps to form few large planets ($\geq M_\oplus$) instead of the hundreds of Mars-sized objects during the formation of planets, as discovered by Levison et al. (2015).

The tandem planet formation is also compatible with the reported compositional distribution of asteroids in asteroid belts (DeMeo and Carry, 2014), including the asteroid belt being composed of a mixture of primitive objects and objects with thermal alternations. In the tandem planet formation, the objects located now in asteroid belts were formed outside the main belt. In other words, they are diffused from the outer MRI front (icy primitive chemical composition) and the inner MRI front (rocky thermally metamorphosed composition). In the later phases, they are scattered into the asteroid belts by the gravitational interaction of large objects, like protoplanets.

Obviously, many subjects remain for future studies. Among them, dispersal of the gas from the disk, formation of chondrules and calcium-aluminum-rich inclusions (CAIs), and planet migration during the latest stages of planet formation are among the most interesting to explore, though they are beyond the scope of this work.

The temperature in the inner most part of the gas disk is as high as 2000K, which is well above the melting temperature of silicate (chondrules) and carcium-alminium rich inclusions (CAIs). The solid particles that melt in the disk are eventually exposed to the radiation field of the central star through photoevapolation during the later phase of the disk evolution. They then interact with stellar wind to get a high outer-bound velocity. As their distances from the central star increase, they cooled down to the melting temperature again. Particles once melted may be recycled in the outer disk to be calcium-aluminum rich inclusions (CAIs) or chondrules.

The last stage of planet formation in the tandem disk must be explored taking into account the various levels of migration processes through both population-synthesis simulations and N-body simulations to explain the varieties of the exosolar planetary systems. In such a later stage, the situation would become much complex because of many factors, such as gravitational interaction among growing planetesimals, depletion of pebbles, the saturation of planetesimal growing due to gravitational heating of planetesimals,

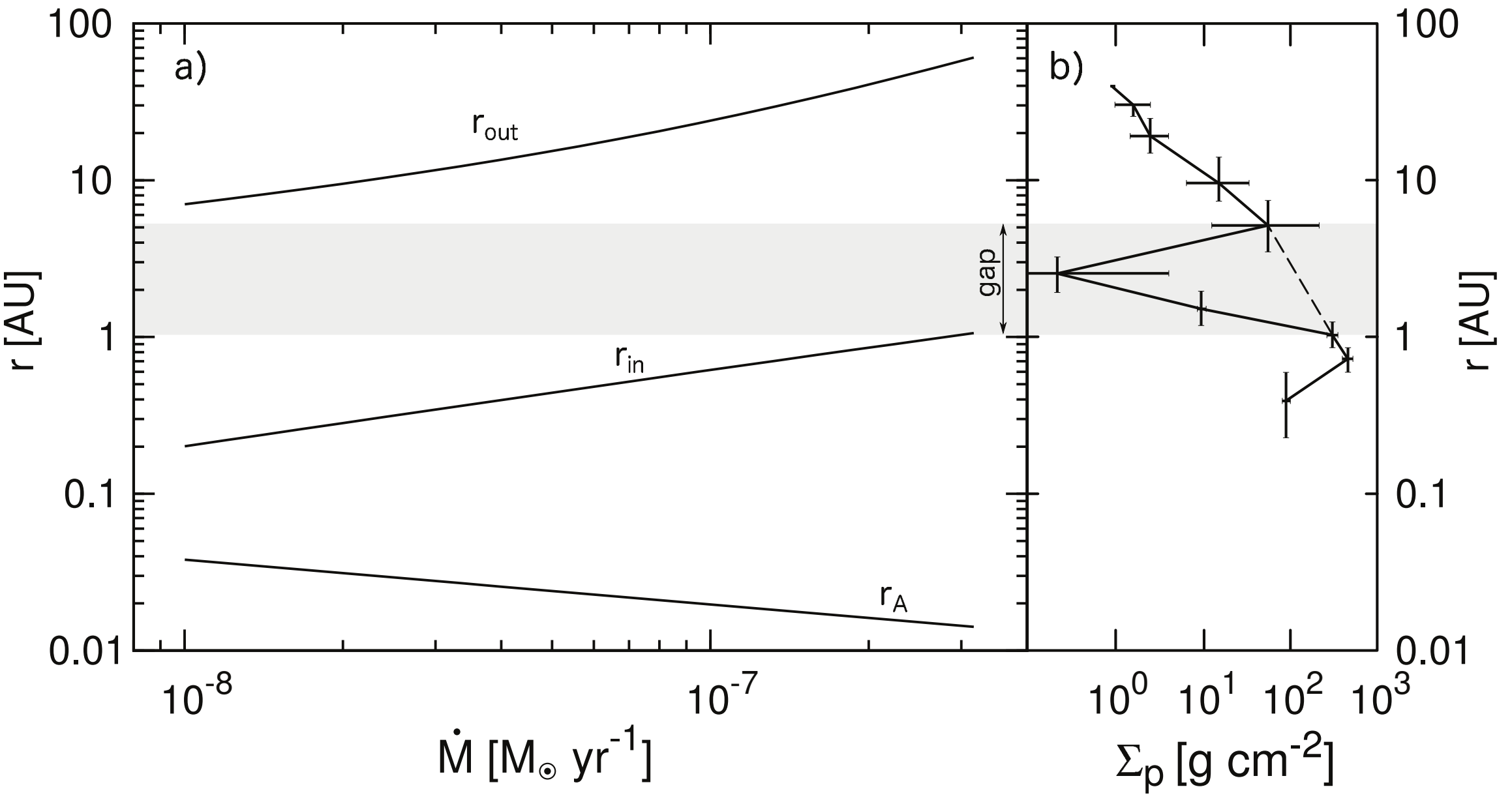

Figure 16: a) The radii of the outer and inner MRI fronts($r_{\rm out}$ and $r_{\rm in}$) and the Alfven radius ($r_{\rm A}$) are plotted against $\dot{M}$. b) The distribution of the solid components in the solar system plotted against the distance from the Sun (Scanned and altered from Weidenschilling, 1977b)

and various types of planetary migrations. We are organizing such studies based on the results of the present work, and will report in the following papers.

We found that this tandem planet formation regime does not take place when magnetic field of the disk considerably weaker (by a factor of 5), since the outer MRI front shifts outward beyond 100 AU. In such a case, most of the solid materials in the 1-100 AU drift down to the inner MRI front to form planetesimals, unlike the case of the tandem regime, described here. We plan to explore what happens in such a case near future. It may explain the variation of planet distributions by the magnetic fields and angular momenta in the parent molecular clouds.

## 6. acknowledgments

We thank Prof. S. Maruyama for his encouraging discussions and Dr. James Dohm for the improvement of English.. This work was partially supported by Grant-in-Aid for Scientific Research on Innovative Areas Grant Number 26106006.